\documentclass[sigplan,screen]{acmart}

\usepackage{pervasives}

\begin{document}

\setcopyright{none}
\acmDOI{}
\acmYear{2018}
\copyrightyear{2018}
\acmISBN{}

\title{Putting in All the Stops: \\ Execution Control for JavaScript}

\author{Samuel Baxter}
\affiliation{\institution{University of Massachusetts Amherst}}
\author{Rachit Nigam}
\affiliation{\institution{University of Massachusetts Amherst}}
\author{Joe Gibbs Politz}
\affiliation{\institution{University of California San Diego}}
\author{Shriram Krishnamurthi}
\affiliation{\institution{Brown University}}
\author{Arjun Guha}
\affiliation{\institution{University of Massachusetts Amherst}}

\lstset{language=JavaScript}

\begin{abstract}
Scores of compilers produce JavaScript, enabling
programmers to use many languages on the Web, reuse existing code, and even
use Web IDEs. Unfortunately, most compilers inherit the
browser's compromised execution model, so
long-running programs freeze the browser tab, infinite loops crash IDEs, and
so on.
The few compilers that avoid these problems suffer poor performance and are difficult to engineer.

This paper presents \system{}, a source-to-source compiler that extends
JavaScript with debugging abstractions and blocking operations, and easily
integrates with existing compilers. We apply \system{} to \numlang{}
programming languages and develop a Web IDE that supports stopping,
single-stepping, breakpointing, and long-running computations. For nine languages, \system{}
requires no or trivial compiler changes. For eight, our IDE is the first that
provides these features. Two of our subject languages have compilers
with similar features. \system{}'s performance is competitive
with these compilers and it makes them dramatically simpler.

\system{}'s abstractions rely on first-class continuations, which it
provides by compiling JavaScript to JavaScript.
We also identify sub-languages of JavaScript that compilers
implicitly use, and exploit these to improve
performance. Finally, \system{} needs to
repeatedly interrupt and resume program execution. We use
a sampling-based technique to estimate program speed that outperforms
other systems.
\end{abstract}

\maketitle
\renewcommand{\shortauthors}{S. Baxter, R. Nigam, J.G. Politz, S. Krishnamurthi, and A. Guha}
\renewcommand{\shorttitle}{Putting in All the Stops: Execution Control for JavaScript}

\section{Programming On the Web}
\label{intro}

Scores of programming languages now compile to JavaScript and
run on the Web~\cite{langs-to-js} and there are several Web IDEs in widespread
use~\cite{whalesong,pyret,burckhardt:touchdevelop,lively,tang:codeskulptor,codecademy-js,codeschool-js,treehouse-js,codio,vocareum}.
This growing audience for Web IDEs and languages that
run in the
browser includes both professionals and students.
Unfortunately, JavaScript and the Web platform lack the abstractions
necessary to build IDEs and serve as a complete compilation target for high-level languages.
As a result, most compilers that produce JavaScript compromise on semantics and most
Web IDEs compromise on basic debugging and usability features.
Furthermore,
as we explain in \cref{relwork}, new technologies such as WebAssembly and
Web Workers do \emph{not} address most of the problems that we
address. Instead, our work may be viewed as presenting
additional challenges to the creators of those technologies.

\paragraph{Limitations in Web IDEs}

The key problem facing a Web IDE is that JavaScript
has a single-threaded execution environment. An IDE that wants to provide a
``stop'' button to halt runaway computation faces a nontrivial
problem, because the callback for that button gets queued for
execution behind the running code---which is not going to
terminate. Related to this, to make pages responsive, the browser
threatens to interrupt computations that run longer than a few
seconds. This means long-running computations need to be broken up
into short events. In turn, because computations cannot run
for very long, JavaScript engines in browsers tend to provide very
shallow stacks, which is problematic for functional programs that rely on recursion.

These limitations are not at all hypothetical.
For example, Codecademy has a
Web IDE for JavaScript and for Python. In response to several
message board requests, Codecademy has a help article that explicitly addresses
infinite loops that freeze the browser~\cite{codecademy-freezing}:
they suggest refreshing the page, which loses browser state and recent
changes. Other systems, such as CodeSchool, Khan Academy, and Python
Tutor~\cite{codechella}, address this problem by killing all programs that run
for more than a few seconds. These problems also
afflict research-driven programming languages, such as
Elm~\cite{czaplicki:elm} and Lively
Kernel~\cite{lively}, which crash when given an infinite loop. \Cref{s:web-ide-bad-eg} discusses all
these systems in more detail.

Some Web IDEs run user code on servers. However, this approach has its own
limitations. The provider has to pay for potentially unbounded server time,
servers must run untrusted code, and state (e.g., time) may reflect the server
and not the client. Moreover, users have to trust servers, cannot work
offline, and cannot leverage the browser's DOM environment.
In this paper, we focus on IDEs that run user code in the browser.

\paragraph{Preliminary Solutions}

There are a handful of robust programming language implementations
 for the Web: 
GopherJS (Go)~\cite{gopherjs}, 
Pyret~\cite{pyret},
Skulpt (Python)~\cite{skulpt}, 
Doppio (JVM)~\cite{vilk:doppio},
GambitJS (Scheme)~\cite{thivierge:gambit-js}, and 
Whalesong (Racket)~\cite{whalesong}.
They use sophisticated compilers and runtime systems to support
some subset of
long-running computations,
shared-memory concurrency, blocking I/O, proper tail calls, debugging,
and other features that are difficult to implement in the browser. However,
these systems have several shortcomings.

First, these systems are difficult to build and maintain because they must
effectively implement expressive control operators that JavaScript does not
natively support.
For example, GopherJS has had several issues in its implementation of
goroutines~\cite{gopherjs-issue-209,gopherjs-issue-225,gopherjs-issue-381,gopherjs-issue-493,gopherjs-issue-426,gopherjs-issue-698};
Skulpt~\cite{skulpt} has had bugs in its debugger~\cite{skulpt-issue-711,skulpt-issue-723};
and the Pyret IDE has problems (\cref{s:pyret}), several of which remain unresolved. 
In fact, the team that
built Pyret previously developed a Web IDE for Scheme~\cite{wescheme}, but
could not reuse the compiler from the Scheme system, because the
execution control techniques and the language's semantics were too tightly coupled.

Second, these systems force all programs
to pay for all features.
For example, Pyret forces all programs to pay the cost of debugging instrumentation,
even if they are running in production or at the command-line; GopherJS forces all programs
to pay the cost of goroutines, even if they don't use concurrency; and Doppio
forces all programs to pay the cost of threads, even if they are single-threaded.
Third, these compilers have a single back-end---one that is presumably
already complex enough---for all browsers,
hence do not maximize performance on any particular browser.
Finally, it is hard for a compiler author to try a new approach,
since small conceptual changes can require the entire compiler and runtime system
to change. Therefore, although these systems
use techniques that are
interchangeable in principle, in practice they cannot share code to benefit
from each others' performance improvements and bug fixes.
What is called for is a clear separation of concerns.

\begin{figure}[t]

\lstset{language=JavaScript}
\begin{lstlisting}
type Opts = {
  cont: 'checked' | 'exceptional' | 'eager', // continuation representation
  ctor: 'direct' | 'wrapped',                    // constructor representation
  timer: 'exact' | 'countdown' | 'approx',       // time estimator
  yieldInterval: number,                         // yield interval
  stacks: 'normal' | 'deep',                     //  deep stacks
  implicits: true | false | '+',                 // implicit conversions
  args: true | false | 'mixed' | 'varargs',      // arity-mismatch behavior
  getters: true | false,                         // support getters
  eval: true |  false                            // apply stopify to eval'd code
}

type AsyncRun = {
  run: (onDone: () => void) => void,
  pause: (onPause: () => void) => void,
  resume: () => void
}

function stopify(source: string, opts: Opts): AsyncRun; #\label{f:stopify-api:fun}#
\end{lstlisting}

\caption{A portion of the \systemlst \textsc{api}.}
\label{f:stopify-api}
\end{figure}

\begin{figure*}[!t]
  \begin{subfigure}[t]{0.23\textwidth}
\begin{tikzpicture}
\node{\pgfimage[width=.95\columnwidth]{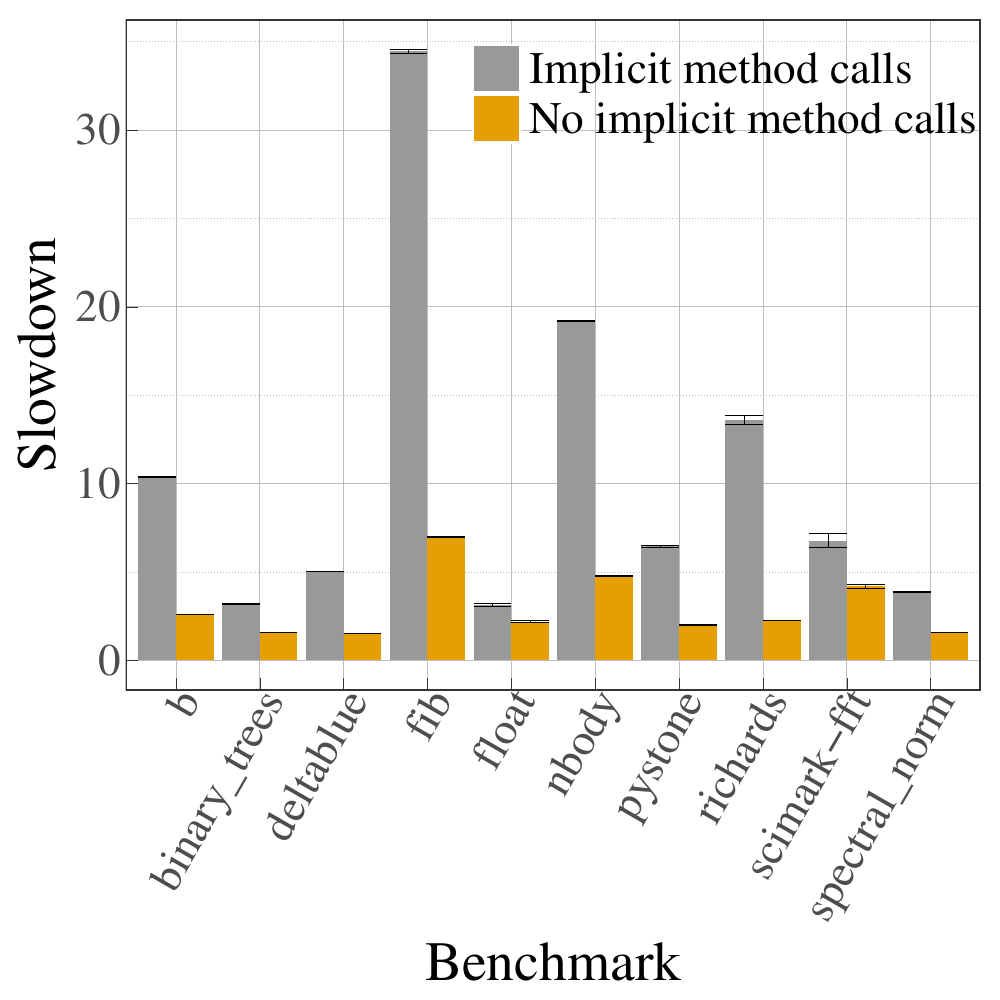}};
\end{tikzpicture}
\caption{Nonterminating arithmetic.}
\label{sane-vs-insane}
\end{subfigure}
\quad
\begin{subfigure}[t]{0.33\textwidth}
  \begin{tikzpicture}
  \node{\pgfimage[width=.95\columnwidth]{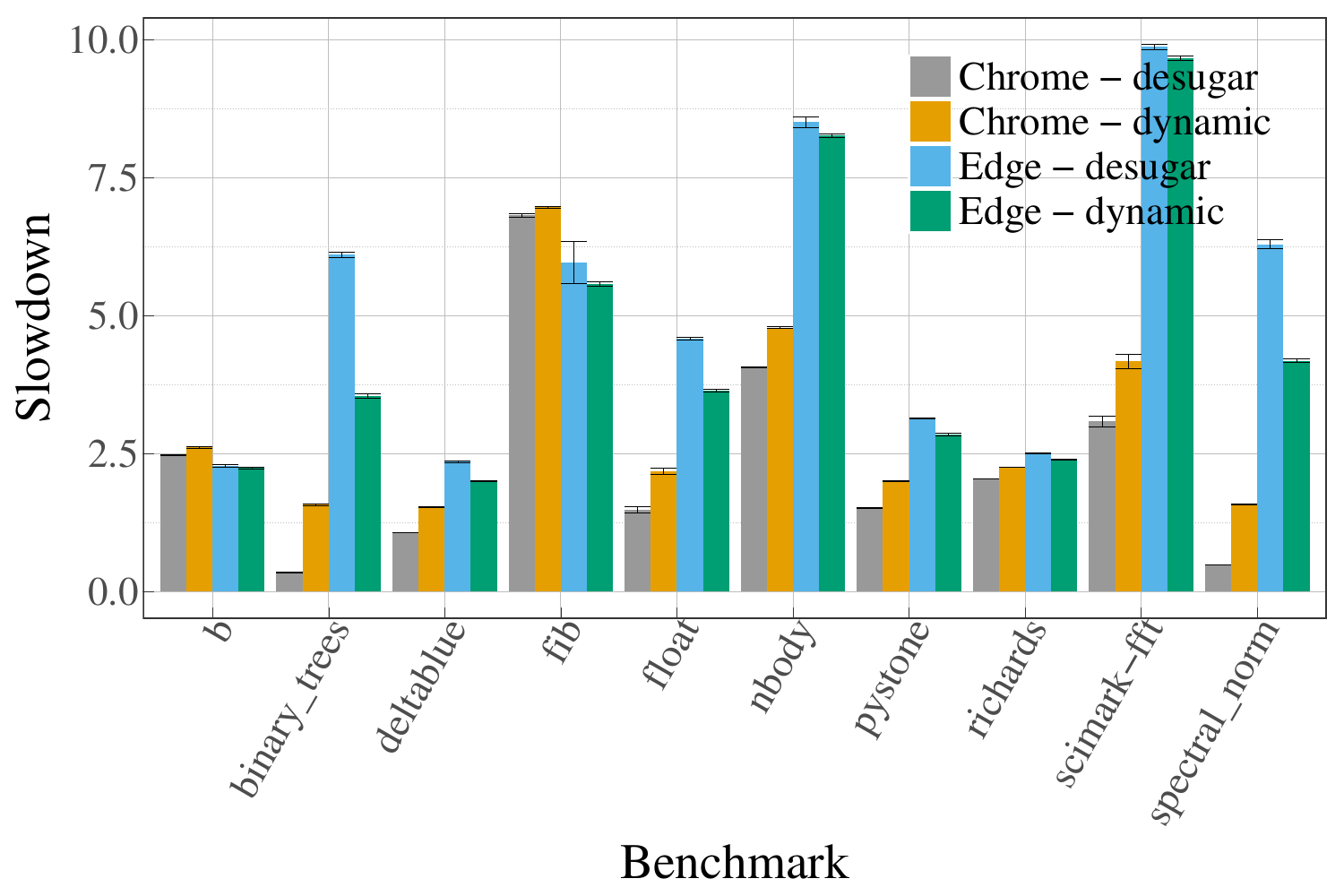}};
  \end{tikzpicture}
  \caption{Constructor encoding.}
  \label{new-method}
\end{subfigure}
\quad
\begin{subfigure}[t]{0.33\textwidth}
  \begin{tikzpicture}
    \node{\pgfimage[width=.95\columnwidth]{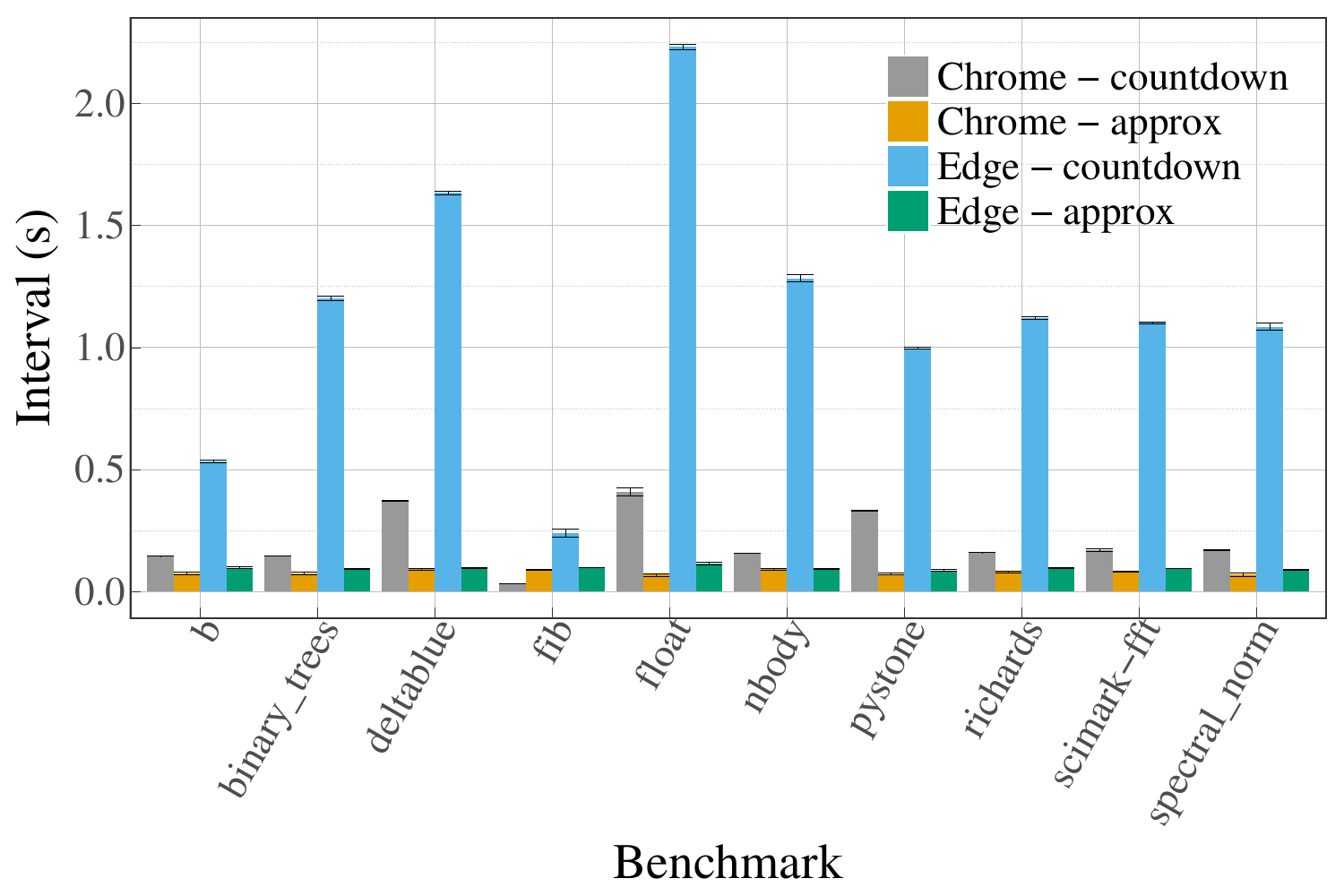}};
    \end{tikzpicture}
    \caption{Average time between yields.}
    \label{estimation-interval}
\end{subfigure}

\caption{Performance of \system{} relative to
unmodified PyJS on a suite of 10 Python benchmarks run 10 times each.
 Each graph shows how an option setting in \system{} affects running
time or latency. Error bars show the 95\% confidence interval.}
\end{figure*}

\paragraph{Our Approach}
Our goal is to enhance JavaScript to make it a suitable target for
Web IDEs and, more broadly, to run a variety of languages atop
JavaScript without compromising their semantics and programming
styles. Our system, \system{}, is a \emph{compiler from JavaScript to
  JavaScript}. Given a na\"{\i}ve compiler from language $L$ to
JavaScript---call it $L$JS---we can compose it with \system{}. \system{}
prepares the code for Web execution while leaving $L$JS
mostly or entirely unchanged.

\system relies on four key ideas. The first is to reify continuations
with a family of implementation strategies
(\cref{sec:callcc}). The second is to identify reasonable
sub-languages of JavaScript---as targeted by compilers---to reduce
overhead and hence improve performance (\cref{js-subsets}). The third
is to dynamically determine the rate at which to yield control to
the browser, improving performance without
hurting responsiveness (\cref{s:long}). Finally, we study how these
different techniques vary in performance across browsers, enabling
browser-specific performance improvements (\cref{sec:evaluation}).

Continuations and execution control features enable new capabilities for
languages that compile to JavaScript. We show several:
(1) \system{} supports long running
computations by periodically yielding control to the browser; (2)
\system{} provides abstractions that help
compilers simulate
blocking operations atop nonblocking APIs; (3) \system{} enables stepping debuggers via source maps; (4) \system{}
allows simulating a much deeper stack than most browsers provide; and (5) \system{} can
simulate tail calls on browsers that don't implement them natively.

We evaluate the effectiveness of \system{} in five ways:
\begin{enumerate}
\item We evaluate \system{} on \numlang{} compilers
that produce JavaScript, nine of which require no
changes, and quantify the cost of \system{} using \numbench{} benchmarks (\cref{main-eval}).

\item We use \system{} to build an
IDE for nine languages. For eight of them, ours is the first Web IDE that
supports long-running computation and graceful termination. For \numstepping{}
languages, our IDE also supports breakpoints and single-stepping (\cref{s:stack-tricks}).

\item We apply \system{} to two well-known JavaScript benchmark suites and examine
  the difference in performance between them (\cref{s:octane-kraken}).

\item We show that our \system{}-based Python IDE is faster and
more reliable than a widely-used alternative (\cref{s:eval-python}).

\item We present a case
study of Pyret, integrating \system{} into its compiler and IDE. We show that
\system{} makes the compiler significantly simpler, helps fix outstanding bugs,
has competitive performance, and presents new opportunities for optimization
(\cref{s:pyret}).

\end{enumerate}

\section{Overview}
\label{sec:overview}

\system, which is written in TypeScript, provides a function called
\systemlst (line~\ref{f:stopify-api:fun} in \cref{f:stopify-api})
that takes two arguments: (1) JavaScript code to run and (2) a set of options
that affect the \system{} compiler and runtime system, which we elucidate in
the rest of this section. The function produces an object with three methods:

\begin{enumerate}

\item The \lstinline|run| method returns immediately and starts to
evaluate the program.
\system{} instruments every function and loop in the program to
periodically save their continuation,
yield control to the browser, and schedule resumption.
In the absence of intervening events, the resumption event runs
immediately (by applying the continuation).
These periodic yields ensure that the browser tab remains
responsive. When execution finally concludes, \system{} applies
the callback passed to \lstinline|run| to notify the caller.

\item The \lstinline|pause| method interrupts the running program by
  setting a runtime flag that \system{} checks before applying a
  continuation in the resumption event. When the flag is set,
  \system{} calls the callback passed to \lstinline|onPause| instead. Therefore,
   in an IDE, the event handler for a ``stop'' button can simply call
\lstinline|pause| and rely on \system{} to handle the rest.

\item Finally, \lstinline|resume| resumes a paused program.

\end{enumerate}

\system{} has additional methods to set breakpoints and simulate blocking
operations, which we elide here.
The rest of this section highlights a few features of \system{}
using, as an example, the
PyJS Python-to-JavaScript compiler's output as \system's input. This section presents only 10
benchmarks running on the Chrome and Edge Web browsers. 
\Cref{sec:evaluation} describes our experimental setup and an extensive evaluation.

\paragraph{Sub-languages of JavaScript}
\label{s:sublangs}

There are some surprising ways to write nonterminating programs in JavaScript.
For, instance, an apparent arithmetic expression like \lstinline|x + 1| or
field access like \lstinline|o.f| can fail to terminate if
the program has an infinite loop in a \lstinline|toString| method or a getter,
respectively. \system{} can
stop these nonterminating programs, but at a significant
performance cost. However, most compilers generate programs in a
``well-behaved'' sub-language of JavaScript.
As \cref{sane-vs-insane} shows, when we compile PyJS with conservative
settings, the slowdown is much higher than when we
specify that arithmetic expressions do not cause infinite loops, which is
a safe assumption to make \emph{of the code generated by PyJS}. \Cref{js-subsets} presents more
sub-languages of this kind.

\paragraph{Browser-specific Optimizations}

Different browsers optimize
JavaScript differently, which gives \system{} the opportunity to
implement browser-specific optimizations. For example, any
function may serve as a constructor and there are two ways to capture a
constructor's continuation frame: we can (1) \emph{desugar} constructor calls
to ordinary function calls (using \lstinline|Object.create|) or (2)
\emph{dynamically} check if a function is a constructor. \Cref{new-method} plots the slowdown incurred by both
approaches on Chrome and Edge. Desugaring is
better than the dynamic approach on Chrome ($p=0.001$), but the other
way around in Edge ($p=0.042$).  To our knowledge, none of the works we
cite in this paper
implement browser-specific optimizations. \Cref{sec:callcc} discusses
compilation issues in more detail.

\paragraph{Estimating Elapsed Time}

\system{} needs to periodically interrupt execution and yield control to the
browser's event loop, which can be done in several ways. Skulpt~\cite{skulpt}
continuously checks the system time, but this is needlessly expensive.
Pyret~\cite{pyret} counts basic blocks and interrupts after executing a fixed
number of them. However, Pyret's approach results in high variability in
page responsiveness between benchmarks and browsers.
\Cref{estimation-interval} shows the average time between interrupts when
counting up to 1 million using this mechanism. On Chrome, a
few benchmarks yield every 30 milliseconds, which is too frequent and
needlessly slows programs down. In contrast, on Edge, some benchmarks only
yield every 1.5 seconds, which is slow enough for the browser to display a
\emph{``script is not responding''} warning.
In \system{},  we sample the system time and dynamically
estimate the rate at which statements are executed. Using this mechanism,
\system{} takes a desired interrupt interval, $\delta$, as a parameter and ensures that
the program interrupts every $\delta$ milliseconds with high probability.
\Cref{estimation-interval} shows that the average time between
interrupts with $\delta\!=\!100$ is very close to $\delta$ on both browsers.

In sum, we've identified the JavaScript sub-language
that PyJS targets, applied browser-specific optimizations, and used
an estimator of system time. This combination makes \system{}'s
Python faster and more reliable than a widely-used alternative (\cref{s:eval-python}),
while also being simpler to maintain. \system{} has several more features and options
that we use to support a variety of programming languages that the rest of this
paper describes in detail.

\section{Continuations for JavaScript}
\label{sec:callcc}

\lstset{language=JavaScript}

This section presents \system{}'s first-class continuations
for a fragment of JavaScript
that excludes some of its most egregious features (which we defer to
\Cref{js-subsets}).

\paragraph{Language Extension}
We extend JavaScript with Sitaram and Felleisen's unary \emph{control}
operator~\cite{sitaram:control-hierarchies}, which we
write as $\kwcontrol$. When
applied to a function, \lstinline|control(function($k$) { $\mathit{body}$ })|,
the operator reifies its continuation to a value, binds it to $k$, and then evaluates $\mathit{body}$ in
an empty continuation. Applying $k$ aborts the current
continuation and restores the saved continuation. For example, the result of
the following expression is \lstinline|0|:
\begin{lstlisting}
10 + control(function (k) { return 0; })
\end{lstlisting}
Above, \kwcontrol{} evaluates the body in an empty
continuation, thus produces \lstinline|0|. In contrast, the next expression
produces \lstinline|11|:
\begin{lstlisting}
10 + control(function (k) { return k(1) + 2; })
\end{lstlisting}
Above, \kwcontrol{} evaluates the body in an empty continuation, then
the application in the body discards its continuation
(which adds \lstinline|2|) and restores the saved continuation (which adds
\lstinline|10|).

\system{} breaks up long running computations using a combination of
\kwcontrol{} and browsers' timers. For example, the following function saves
its continuation and uses \lstinline|setTimeout| to schedule an event that
restores the saved continuation:
\begin{lstlisting}
function suspend() {
  control(function(k) { window.setTimeout(k, 0); }); }
\end{lstlisting}
Therefore, \lstinline|suspend()| gives the browser a chance to
process other queued events
before resuming the computation. For example, the following infinite loop does
not lock up the browser since each iteration occurs in a separate event:
\begin{lstlisting}
while(true) { suspend(); }
\end{lstlisting}
We now discuss how we compile \lstinline|control| to ordinary
JavaScript.

\paragraph{Strawman Solutions}
There are two natural ways to implement \kwcontrol{} in modern JavaScript. One
approach is to transform the program to continuation-passing style
(\textsc{cps})
(with trampolines if necessary). Alternatively, we could use generators
to implement one-shot continuations~\cite{call1cc}. We
investigated and abandoned both approaches for two reasons. First,
on several benchmarks, we find that \textsc{cps} and generators are 3x and 2x
slower than the approach we present below. Second, both approaches change
the type of instrumented functions (\textsc{cps} adds an extra argument
and generators turn all functions into generator objects). This makes it
hard to support features such as constructors and prototype inheritance.

\begin{figure}
\lstset{language=JavaScript}

\begin{lstlisting}
var mode = 'normal';
var stack = [];

function P(f, g, x) {
  var t0, $\ell$, $k$;
  if (mode === 'restore') { #\label{restore-begin}#
    $k$ = stack.pop();
    [t0] = $k$.locals; #\label{restore-locals}#
    $\ell$ = $k$.label;
    $k$ = stack[stack.length - 1];
  } #\label{restore-end}#
  function locals() { return [t0]; }
  function reenter() { return P(f,g,x); }
  if (mode === 'normal' || (mode === 'restore' && $\ell$ === 0)) { #\label{P-check-g}#
    t0 = mode === 'normal' ? g(x) : $k$.reenter(); #\label{call-g}#
    if (mode === 'capture') { #\label{capture-g}#
      stack.push({ label: 0, locals: locals(), reenter: reenter }); #\label{push-g}#
      return; #\label{return-g}#
    }
  }
  return f(t0); #\label{call-f}#
}
\end{lstlisting}
\caption{An instrumented function.}
\label{compose3-retval}
\end{figure}

\begin{figure*}
\begin{minipage}{\textwidth}
\footnotesize
\(
\begin{array}{@{}r@{\;}c@{\;}l}
\Jumper{x~\jsop{=}~e} & =
  & \jskw{if}~(\mathit{normal})~\jsop{\{}\;x\jsop{=}e\;\jsop{\}~} \\
\Jumper{s_1\jsop{;}~s_2} & =
  & \Jumper{s_1}\jsop{;}~\Jumper{s_2} \\
\Jumper{x~\jsop{=}~f_j\jsop{(}e_1\cdots e_n\jsop{)}} & =
  & \jskw{if}~\jsop{(}\mathit{normal}~\jsop{||}~\ell~\jsop{===}~j\jsop{)}~
    \jsop{\{}\;\Apply{x,f_j,e_1,\cdots,e_n}\;\jsop{\}} \\
\Jumper{\jskw{if}~\jsop{(}e\jsop{)}~s_1~\jskw{else}~s_2} & =
  & \jskw{if}~\jsop{((}\mathit{normal}~\jsop{\&\&}~e\jsop{)}~\jsop{||}~
    \jsop{(}\mathit{restore}~\jsop{\&\&}~\ell\in s_1\jsop{))}~\Jumper{s_1}~
    \jskw{else}~\jskw{if}~\jsop{(}\mathit{normal}~\jsop{||}~
      \jsop{(}\mathit{restore}~\jsop{\&\&}~\ell\in s_2\jsop{))}~\Jumper{s_2} \\
\Jumper{\jskw{function}~f\jsop{(}x_1\cdots x_n\jsop{)}~\jsop{\{}~\jskw{var}~
  y_1\cdots y_m\jsop{;}~s~\jsop{\}}} & =
  & \jskw{function}~f\jsop{(}x_1\cdots x_n\jsop{)}~\jsop{\{}~\jskw{var}~
    y_1\cdots y_m\jsop{,}\;\ell\jsop{,}\;k\jsop{;}\;\mathit{restoreFrame}
    \jsop{;}\;\mathit{locals}\jsop{;}\;\mathit{reenter}\jsop{;}\;\Jumper{s}~
    \jsop{\}}  \\
\mathit{normal} & \triangleq &
  \jsop{mode}~\jsop{===}~\jsop{'normal'} \\
\mathit{restore} & \triangleq &
  \jsop{mode}~\jsop{===}~\jsop{'restore'} \\
\mathit{capture} & \triangleq &
  \jsop{mode}~\jsop{===}~\jsop{'capture'} \\
\mathit{restoreFrame} & \triangleq
  & \jskw{if}~\jsop{(}\mathit{restore}\jsop{)}~\jsop{\{}~k~\jsop{=}~
    \jsop{stack.pop();}\;\jsop{[}y_1\cdots y_m\jsop{]}\;\jsop{=}\;
    k\jsop{.locals()}\jsop{;}~k\;\jsop{=}\;k\jsop{.last}\jsop{;}~\jsop{\}} \\
\mathit{locals} & \triangleq
  & \jskw{var}~\jsop{locals}~\jsop{=}~\jsop{()}~\jsop{=>}~
    \jsop{[}y_1 \cdots y_m\jsop{]} \\
\mathit{reenter} & \triangleq
  & \jskw{var}~\jsop{reenter}~\jsop{=}~\jsop{()}~\jsop{=>}~
    f\jsop{.call(}\jskw{this}\jsop{,}x_1\cdots x_n\jsop{)} 
\end{array}
\)
\subcaption{Compiling JavaScript to support continuations.}
\label{f:jumper}
\end{minipage}

\begin{minipage}{\textwidth}
\footnotesize
\(
\begin{array}{@{}r@{\;}c@{\;}l}
\Apply{x,f_j,e_1,\cdots,e_n} & =
  & x\;\jsop{=}\;\mathit{normal}~\jsop{?}~
    f\jsop{(}e_1\cdots e_n\jsop{)}~\jsop{:}~k\jsop{.reenter();}~ \\
& & \jskw{if}\;\jsop{(}\mathit{capture}\jsop{)}\;\jsop{\{}\;k\jsop{.push(\{}~
    \jsop{label:}\;j\jsop{,}\;\jsop{locals:}\;\jsop{locals(),}\;
    \jsop{reenter}\;\jsop{\});}\;\jskw{return}\jsop{;}\;\jsop{\}} \\
\kwcontrol & \triangleq
  & \jskw{function}\;\jsop{(f)}\;\jsop{\{}\;\jsop{stack.push(\{}\;
    \jsop{reenter:}\;\jsop{f}\;\jsop{\});}\;\jsop{mode}\;\jsop{=}\;\jsop{'capture'}\;
    \jsop{\}}
\end{array}
\)
\subcaption{Checked-return continuations.}
\label{f:jumper-return}
\end{minipage}

\begin{minipage}{\textwidth}
\footnotesize
\(
\begin{array}{@{}r@{\;}c@{\;}l}
\Apply{x,f_j,e_1,\cdots,e_n} & =
  & \jskw{try}\;\jsop{\{}x\;\jsop{=}\;\mathit{normal}~\jsop{?}~
    f\jsop{(}e_1\cdots e_n\jsop{)}~\jsop{:}~k\jsop{.reenter();}\;\jsop{\}} \\
& & \jskw{catch}\;\jsop{(exn)}\;\jsop{\{}\;
    \jskw{if}\;\jsop{(}\mathit{capture}\jsop{)}\;\jsop{\{}\;k\jsop{.push(\{}~
    \jsop{label:}\;j\jsop{,}\;\jsop{locals:}\;\jsop{locals(),}\;
    \jsop{reenter}\;\jsop{\});}\;\jskw{throw}\;\jsop{exn;}\;\jsop{\}\;\}} \\
\kwcontrol & \triangleq
  & \jskw{function}\;\jsop{(f)}\;\jsop{\{}\;\jsop{stack.push(\{}\;
    \jsop{reenter:}\;\jsop{f}\;\jsop{\});}\;\jsop{mode}\;\jsop{=}\;
    \jsop{'capture';}\;\jskw{throw}\;\jsop{'capture';}\;\jsop{\}}
\end{array}
\)
\subcaption{Exceptional continuations.}
\label{f:jumper-exn}
\end{minipage}

\begin{minipage}{\textwidth}
\footnotesize
\(
\begin{array}{@{}r@{\;}c@{\;}l}
\Apply{x,f_j,e_1,\cdots,e_n} & =
  & k\jsop{.push(\{}~
    \jsop{label:}\;j\jsop{,}\;\jsop{locals:}\;\jsop{locals(),}\;
    \jsop{reenter}\;\jsop{\});} \\
& & x\;\jsop{=}\;\mathit{normal}~\jsop{?}~
    f\jsop{(}e_1\cdots e_n\jsop{)}~\jsop{:}~k\jsop{.reenter();}~ \\
\kwcontrol & \triangleq
  & \jskw{function}\;\jsop{(f)}\;\jsop{\{}\;\jsop{stack.push(\{}\;
    \jsop{reenter:}\;\jsop{f}\;\jsop{\});}\;\jsop{mode}\;\jsop{=}\;
    \jsop{'capture';}\;\jskw{throw}\;\jsop{'capture';}\;\jsop{\}}
\end{array}
\)
\subcaption{Eager continuations.}
\label{f:jumper-eager}
\end{minipage}

\caption{Compiling the \kwcontrol{} operator to ordinary JavaScript.}

\end{figure*}
\subsection{Our Approach}
\label{s:approach}

We compile \lstinline|control| to ordinary JavaScript in three steps: (1)
A-normalize~\cite{anf} programs, thus translating to a JavaScript subset where all
applications are either in tail position or name their result; (2)
box assignable variables that are captured by nested functions (discussed in \cref{assignable-vars}); (3)
instrument every function to operate in three modes: in
\emph{normal mode}, the program executes normally; in \emph{capture mode}
the program unwinds and reifies its stack; and in \emph{restore mode} the
program restores a saved stack.
We run the program in the context of a driver loop that manages the transition
between these modes.

We use the following function \lstinline|P| as a running example:
\begin{lstlisting}
function P(f, g, x) {  return f(g(x)); }
\end{lstlisting}
Although \lstinline|P| does not use \lstinline|control| itself, the
functions that it applies may use \lstinline|control|.
\Cref{compose3-retval} shows an instrumented version of \lstinline|P| and two
global variables that determine the current execution mode (\lstinline|mode|)
and hold a reified stack (\lstinline|stack|).
In normal mode, the instrumented function is equivalent to the original function
and the reified stack is not relevant.

Suppose \lstinline|g(x)| applies \lstinline|control|, which switches
execution to capture mode.
To address this case, \lstinline|P| checks to see if the program is in capture
mode  after \lstinline|g(x)| returns (line~\ref{capture-g}).
If so, \lstinline|P|
reifies its stack frame (line~\ref{push-g}) and
returns immediately.
A reified stack frame contains (1) a copy of the local variables,
(2) a label that identifies
the current position within the function, and (3) a thunk called \lstinline|reenter|
that re-enters the function.

Now, consider how \lstinline|P| operates in restore mode.
The function begins with a block of code that restores
the saved local variables (lines \ref{restore-begin}---\ref{restore-end}).
Next, on line~\ref{P-check-g}, the function checks the saved
label to see if \lstinline|g(x)| should be applied. (The  only label
in this function is \lstinline|0|.) Finally, instead of applying
\lstinline|g(x)| again, we apply the \lstinline|reenter()| function that
\lstinline|g(x)| had pushed onto the stack during capture mode.
When \lstinline|g(x)| returns normally, the last line of the function calculates
\lstinline|f(t0)|, where \lstinline|t0| is the result of \lstinline|g(x)|.

In general, to implement the three execution modes, we transform a function
\lstinline|f| as follows. (1) We define a nested thunk called
\lstinline|locals| that produces an array containing the values of
\lstinline|f|'s local variables. (2) We define a nested thunk called
\lstinline|reenter| that applies \lstinline|f| to its original arguments. (3)
We give a unique label to every non-tail call in \lstinline|f|. (4) After every
non-tail call, we check if the program is in capture mode. If so, we push an
object onto \lstinline|stack| that contains the (a) label of the function call,
(b) an array of local variable values (produced by \lstinline|locals|), and (c)
a reference to the \lstinline|reenter| function. This object is the
continuation frame for \lstinline|f|. We then immediately return and allow
\lstinline|f|'s caller to do the same. (5) At the top of \lstinline|f|, we add
a block of code to check if the program is in restore mode, which indicates
that a continuation is being applied. If so, we use the reified stack frame to
restore local variables and the label saved on the stack. (6) We instrument
\lstinline|f| such that in restore mode, the function effectively jumps to the
call site that captured the continuation. Here, we invoke the
\lstinline|reenter| method of the next stack frame. When the continuation is
fully restored, execution switches back to normal mode. Finally, the top-level
of the program needs to be wrapped by a thunk and executed within a driver
loop that manages execution in two ways. (1) The expression \lstinline|control(f)|
switches execution to capture mode and reifies the stack.
Therefore, the driver loop has to apply \lstinline|f| to the reified
stack. (2) When a continuation is applied, the program throws a special
exception to unwind the current stack. The driver loop has to start restoring
the saved stack by invoking the \lstinline|reenter| method of the bottommost frame.

The function $\mathcal{K}$ in \Cref{f:jumper} presents
the compilation algorithm for a representative fragment of JavaScript.
The algorithm assumes that expressions ($e$) do not contain function declarations
or applications and that each non-tail function application is subscripted with
a unique label. We write $\ell \in s$ to test if $s$ contains an application
subscripted with $\ell$. We present the rule for capturing stack frames
(the $\mathcal{A}$ function) and the definition of \kwcontrol{}
in a separate figure (\Cref{f:jumper-return}),
because \system{} can capture continuations in other ways
(\cref{cont-variations}).

\subsubsection{Exception Handlers and Finalizers}

Suppose a program captures a continuation within a
\lstinline|catch| clause. In restore mode, the only way to re-enter
the \lstinline|catch| clause is for the associated \lstinline|try| block to
throw an exception. Throwing an arbitrary exception will lose the
original exception value, so we need to throw the same exception that was
caught before. Therefore, we instrument each \lstinline|catch| block to create
a new local variable that stores the caught exception value and instrument
each \lstinline|try| block to throw the saved exception in restore mode.

Now, suppose a
program captures a continuation within a \lstinline|finally| clause.
In restore mode, the only way to re-enter
the \lstinline|finally| block is to \lstinline|return| within its associated \lstinline|try| block.
However, the returned value is not available within the \lstinline|finally|
block. Therefore, we instrument \lstinline|try| blocks with finalizers to save
their returned value in a local variable. In restore mode, the \lstinline|try|
block returns the saved value to transfer control to the \lstinline|finally|
block, which preserves the returned value.

\subsection{Variations of Our Approach}
\label{cont-variations}

The above approach is parameterizable in two major ways: (1)
how stack frames are captures and (2) how continuations are captured within
constructors.

\paragraph{Capturing Stack Frames}

\system{} can capture stack frames in three different ways. The previous
section presented a new approach that we call \emph{checked-return continuations}:
every function application is instrumented to check whether the
program is in capture mode. \system{} also supports two more approaches
from the literature.
An alternative that involves fewer checks is the \emph{exceptional continuations}
approach~\cite{loitsch:exceptional-continuations}. Here,
\lstinline|control| throws a special exception and every function
application is guarded by an exception handler that catches the special
exception, reifies the stack frame, and re-throws it. Although this approach
involves fewer conditionals than the checked return approach, exception
handlers come at a cost.
Both checked-return and exceptional continuations reify the stack lazily.
An alternative approach, which we call \emph{eager continuations}, maintains
a shadow stack at all times~\cite{thivierge:gambit-js}. This makes capturing continuations trivial
and very fast. However, maintaining the shadow stack slows down normal
execution. A key feature of \system{} is that it unifies
these three approaches and allows them to freely compose with all the other
configuration options that it provides.

\paragraph{Constructors}
JavaScript allows almost any function to be used as a constructor and
a single function may assume both roles.
\system{}  allows constructors to
capture their continuation using two different approaches.
The simple approach is to desugar \lstinline|new|-expressions to ordinary
function calls (using \lstinline|Object.create|), which effectively
eliminates constructors. (Constructors for builtin types, such as \lstinline|new Date()|, cannot be eliminated in this way.)
Unfortunately, this desugaring can perform poorly on some browsers.

It is more challenging to preserve \lstinline|new|-expressions. Consider
the following constructor, which creates two fields and calls a function
\lstinline|f|:
\begin{lstlisting}
function N(x, f) {this.x = x; this.y = f(); return 0;}
\end{lstlisting}
We need to instrument \lstinline|N| to address the case where \lstinline|f|
captures its continuation. The problem is that \lstinline|new N(x, f)| allocates
a new object (bound to \lstinline|this|) every time it is called. Therefore, in restore mode, we have
to apply \lstinline|N| to the originally allocated object so that we do not
lose the write to \lstinline|this.x|. We address this problem in
the \lstinline|reenter| function, by calling \lstinline|N| as a function (not a constructor)
and passing the original value of \lstinline|this|:
\begin{lstlisting}
N.call(this, x, f)
\end{lstlisting}
This presents another problem---when \lstinline|N| is invoked as a constructor 
(as it originally was) it returns \lstinline|this|, but when \lstinline|N|
is applied as an ordinary function (during restoration), it returns \lstinline|0|.
Therefore, we also need to track if the function was originally
invoked as a constructor, so that we can return the right value in
restore mode. We accomplish this by
using  \lstinline|new.target| (ES6)
to distinguish function calls from constructor calls.

\subsubsection{Assignable Variables}
\label{assignable-vars}

A problem that arises with this approach involves assignable
variables that are captured by nested functions. To
restore a function \lstinline|f|, we have to reapply \lstinline|f|,
which allocates new local
variables, and thus we restore local variables' values (e.g.,
line~\ref{restore-locals} in \cref{compose3-retval}). However,
suppose \lstinline|f| contains a nested function \lstinline|g|
that closes over a variable \lstinline|x| that is local to \lstinline|f|.
If \lstinline|x| is an assignable variable, we must ensure
that after restoration
\lstinline|g| closes over the new copy of \lstinline|x| too.
We resolve this problem by boxing assignable variables that are captured
by nested functions. This is the solution that \emph{scheme2js} uses~\cite{loitsch:exceptional-continuations}.

\subsubsection{Proper Tail Calls}

Our approach preserves proper tail calls.
Notice that the application of \lstinline|f| is in tail position and is not 
instrumented (line~\ref{call-f} of \cref{compose3-retval}). Consider what happens if \lstinline|f| uses
\lstinline|control|: \lstinline|f| would first reify its own stack frame and then return immediately to
\lstinline|P|'s caller (instead of returning to \lstinline|P|). In restore
mode, \lstinline|P|'s caller will jump to the label that called \lstinline|P|
and then call \lstinline|nextFrame.reenter()|. Since \lstinline|P| did not save
its own stack frame, this method call would jump into \lstinline|f|, thus preserving proper tail calls.
On browsers that do not support proper tail calls, \system{} uses
trampolines.

\section{Sub-Languages of JavaScript}
\label{js-subsets}

\begin{figure}
\footnotesize
\begin{tabular}{|l|c|c|c|c|l|}
\hline
\textbf{Compiler} & \textbf{Impl} & \textbf{Args} & \textbf{Getters} & \textbf{Eval} & \textbf{(\#) Benchmarks} \\
\hline
PyJS           & \xmark        & \textbf{M} & \xmark     & \xmark     & (16) \cite{pypy-benchmarks,benchmarks-game} \\
ScalaJS        & \lstinline|+| & \xmark     & \xmark     & \xmark     & (18) \cite{benchmarks-game} \\
scheme2js      & \xmark        & \textbf{V} & \xmark     & \xmark     & (13) \cite{larceny-benchmarks-2009} \\
ClojureScript  & \lstinline|+| & \textbf{M} & \xmark     & \xmark     & (8) \cite{benchmarks-game} \\
dart2js        & \lstinline|+| & \xmark     & \textbf{T} & \textbf{T} & (15) \cite{dart-ton80,benchmarks-game} \\
Emscripten     & \xmark        & \textbf{V} & \xmark     & \xmark     & (13) \cite{jetstream,benchmarks-game} \\
BuckleScript   & \xmark        & \xmark     & \xmark     & \xmark     & (15) \cite{operf-micro,benchmarks-game} \\
JSweet         & \lstinline|+| & \textbf{M} & \xmark     & \xmark     & (9) \cite{scimark,benchmarks-game} \\
JavaScript     & \cmark        & \cmark     & \cmark     & \cmark     & (19) \cite{moz-kraken,benchmarks-game} \\
Pyret          & \xmark        & \xmark     & \xmark     & \textbf{T} & (21) \cite{pyret}   \\
\hline
\end{tabular}
\caption{Compilers, their sub-language of JavaScript, and benchmark sources.
A \cmark{} or \xmark{} indicates that a JavaScript feature is used in full or
completely unused. The  other symbols indicate
restricted variants of the feature (discussed in \cref{js-subsets}).}
\label{compiler-configs}
\end{figure}

\system{}'s continuations work with arbitrary JavaScript
(ECMAScript 5) code, but this can come
at a significant cost. Fortunately,
compilers do not generate arbitrary code and every compiler we've studied only
generates code in a restricted
\emph{sub-language of JavaScript}. This section identifies sub-languages
that compilers (implicitly) use. In fact, in several cases, we find that multiple, independently
developed compilers use the same sub-language of JavaScript.
\system{} makes
these sub-languages explicit: the \systemlst function (\cref{f:stopify-api})
consumes a program $p$ and a specification of $p$'s sub-language and then
exploits properties of the sub-language to produce simpler and faster code.

We classify each sub-language as a composition of four orthogonal JavaScript
language features. Each feature is either completely unused (\xmark), used to its
fullest extend (\cmark), or used in a restricted manner. \Cref{compiler-configs}
summarizes the sub-languages that our ten compilers inhabit. Note
that the only language that requires all JavaScript features is JavaScript
itself! A compiler author can always use \system{} with all features enabled.
But, targeting a sub-language of JavaScript will improve performance dramatically.

\subsection{Implicit Operations}
\label{s:impl-op}

In JavaScript, an apparent arithmetic expression like \lstinline|x - 1|
may run forever. This occurs when
 \lstinline|x| is bound to an object that defines a \lstinline|valueOf|
or \lstinline|toString| method. The \lstinline|-| operator implicitly invokes
these methods when \lstinline|x| is an object and these methods may not terminate.
For completeness, \system{} supports all implicit operations, but they are
expensive (\cref{sane-vs-insane}). Fortunately, the only language
that requires all implicits is JavaScript itself (the \cmark{} in the \textbf{Impl}
column of \cref{compiler-configs}).

\paragraph{No Implicits} 
Several compilers do not need JavaScript to make any implicit
calls (\xmark{} in the \textbf{Impl} column).
For these compilers, \system{} can safely assume that all arithmetic expressions
terminate.

\paragraph{Concatenation Only}

In JavaScript, the \lstinline|+| operator is overloaded to perform addition and
string concatenation, and some compilers rely on \lstinline|+| to invoke
\lstinline|toString| to concatenate strings (\lstinline|+| in the \textbf{Impl} column).
For example, the JSweet Java compiler relies on this behavior, since
Java overloads \lstinline|+| and implicitly invokes \lstinline|toString|
in a similar way. 
For these compilers, \system{}
desugars the \lstinline|+| operator to expose implicit calls to
\lstinline|toString| and assumes that other operators do not invoke implicit
methods.

\subsection{Arity Mismatch Behavior}
\label{s:args-object}

JavaScript has no arity-mismatch errors: any function
may receive more arguments (or fewer) than it declares in its formal argument
list. When a function receives fewer arguments, the elided arguments are set to
the default value \lstinline|undefined|. All the arguments, including extra ones,
are available as properties of a special \lstinline|arguments|
object, which is an implicit, array-like object that every function receives.
Some compilers do not leverage this behavior at all (\xmark{} in the \textbf{Args}
column).
\system{} has full support for
\lstinline|arguments|, but we also identify two restricted variants that
compilers use in practice.

\paragraph{Variable Arity Functions} Many compilers use \lstinline|arguments|
to simulate variable-arity functions (\textbf{V} in the \textbf{Args} column).
To restore the continuation frame of a variable-arity function,
\system{} applies it to its \lstinline|arguments| object
instead of explicitly applying it to its formal
arguments:
\begin{lstlisting}
f.apply(this, arguments)
\end{lstlisting}
However, this simple approach breaks down when \lstinline|arguments| is used in other
ways.

\paragraph{Optional Arguments}

A different problem arises when \lstinline|arguments|
simulates optional arguments. Suppose a function \lstinline|f| has formals
\lstinline|x| and \lstinline|y| and that both are optional. If so, \lstinline|f|
can use \lstinline|arguments.length| to check if it received fewer
arguments than it declared and then initialize the missing arguments to default
values. However, this does not affect \lstinline|arguments.length|. Therefore,
if we restore \lstinline|f| by applying it to its \lstinline|arguments| object
(i.e., like a variable-arity function), then the default values will be lost.
So, we need to restore \lstinline|f| by explicitly applying it to its
formal arguments, i.e., \lstinline|f.call(this, x, y)|. This is how
we restore ordinary functions (\cref{s:approach}), thus we don't need a sub-language
when the program uses optional arguments.

\paragraph{Mixing Optional and Variable Arity}

We've seen that variable-arity functions and functions with optional
arguments need to be restored in two different ways. However, we also
need to tackle functions that mix both features (\textbf{M} in the \textbf{Args} column).  To restore these
functions, we effectively apply both approaches simultaneously: we pass
formal parameters explicitly and the \lstinline|arguments| object as a field
on the reified continuation frame.

\paragraph{Complete Support for Arguments}
However, even this does not cover the full
range of possible behaviors. For example, JavaScript allows
formal arguments to be aliased with fields in the
\lstinline|arguments| array. Therefore, if a function updates the same location
as a formal argument and as an element of the \lstinline|arguments| object
then the approach above will break. \system{} supports this
behavior by transforming all formal argument references into
\lstinline|arguments| object indexing.
This comes at a higher cost to performance and is only
necessary when the source language is JavaScript.

\subsection{Getters, Setters, and Eval}
\label{s:getters-setters}

In principle, any read or write to a field may invoke a getter
or a setter that may have an infinite loop.
Fortunately, this issue does not arise for most source languages (\xmark{} in the \textbf{Getters} column).
For example, Scala and Python support getters and setters, but ScalaJS
and PyJS do not use JavaScript's getters and setters to implement them.
On the other hand, Dart does make use of getters and setters but only calls
trivial internal functions that terminate (\textbf{T} in the \textbf{Getters} column).
Therefore, we can safely omit instrumentation for getters for Dart.
If a compiler generates JavaScript with getters and setters, \system{}
can instrument all field reads and writes to capture continuations. However,
\system{} also supports a simple annotation (written as a JavaScript comment)
that indicates that an expression may trigger a getter or a setter. Therefore,
a compiler can be modified to produce this annotation where necessary, which
avoids the cost of instrumenting every field access in the program, which can
be prohibitively expensive.

\system{} supports \lstinline|eval| by rewriting occurrences of \lstinline|eval|
to invoke the \system{} compiler, which is straightforward since it is written
in TypeScript.  However, the compiler and its dependencies are nearly 5MB
of JavaScript and takes much longer than
the browser's native \lstinline|eval| function. Aside from interleaving the
\system{} compiler with program execution, supporting \lstinline|eval| requires
heavier instrumentation that further degrades performance. For instance, all
variables in scope at the time of \lstinline|eval| must be conservatively boxed
in case a variable escapes in the context of evaluating the string.
Moreover, \system{} only supports strict \lstinline|eval|;
non-strict \lstinline|eval| may introduce new variable bindings which can be
referenced by the outer scope, and \system{} cannot distinguish between free
variables and such cases.
Fortunately, most languages do not require JavaScript's \lstinline|eval| (\xmark{} in the \textbf{Eval} column).
(In fact, compilers tend not to support \lstinline|eval|
even when the source language requires it.) Dart and Pyret are two exceptions:
they use \lstinline|eval| as a form of compression: they  dynamically generate lots of trivial functions
(e.g., value constructors) that very obviously terminate (\textbf{T} in the \textbf{Eval} column). 
In these cases, it makes sense to leave \lstinline|eval| uninstrumented.
Finally, we note that the best way to support \lstinline|eval| in the source
language is to lightly modify the source language compiler to pass the
\systemlst function an AST instead of a string. (\system{} uses
a standard representation of JavaScript ASTs~\cite{babylon}.) This avoids needlessly
parsing and regenerating JavaScript at runtime.

\section{Execution Control}
\label{s:long}

We can now say more about the options that \systemlst
(\cref{f:stopify-api}) takes along with the program $p$ to compile: (1) the implementation strategy for continuations
(\cref{s:approach}), (2) the sub-language that $p$ inhabits (\cref{js-subsets}), (3)
whether breakpoints and single-stepping are desired, and (4) whether
$p$ requires a deep stack. \system{} transforms $p$ into an equivalent
program that (1) runs without freezing the browser tab, (2) can be gracefully
terminated at any time, (3) can simulate blocking operations, and (4)
optionally supports deep stacks, breakpoints, and single-stepping.

\begin{figure}
\begin{lstlisting}
var distance = 0, counter = 0, ticks = 0, lastTime = 0, velocity = 0;

function resetTime() {  distance = 0; }

function estimateElapsed() {
  distance++;
  if (counter-- === 0) {
    const now = Date.now();
    velocity = ticks / (now - lastTime); lastTime = now; ticks = $t$ * velocity;
    counter = ticks;
  }
  return distance / velocity;
}

// Stopify's implementation uses postMessage, which is faster than setTimeout.
function defer(k) { setTimeout(k, 0); } // enqueues k() in the event queue

var mustPause = false, saved;
function maySuspend() {
  if (estimateElapsed() >= $\delta$) {
    resetTime();
    control(function (k) {
      return defer(function() {
        if (mustPause) { saved = k; onPause(); } else { k(); }
      }); }); } }

\end{lstlisting}
\caption{\system{} inserts \lstinline|maySuspend()| into programs to support long computations.}
\label{f:suspend}
\end{figure}

\subsection{Long Computations and Graceful Termination}
\label{s:sampling}

\begin{figure}
\centering
\begingroup\small
\scalebox{0.6}{
\begin{tabular}{|l|r|r|r|}
  \hline
Benchmark & Countdown ($\mu \pm \sigma$) & Approximate ($\mu \pm \sigma$) & Exact ($\mu \pm \sigma$) \\
  \hline
  b & 122.4 $\pm$ 27.35 ms & 87.75 $\pm$ 39.33 ms & 114.3 $\pm$ 26.19 ms \\ 
  binary-trees & 146.3 $\pm$ 32.55 ms & 89.18 $\pm$ 24.71 ms & 108 $\pm$ 3.92 ms \\ 
  deltablue & 386.4 $\pm$ 85.99 ms & 97.31 $\pm$ 21.06 ms & 109.2 $\pm$ 1.359 ms \\ 
  fib & 67.63 $\pm$ 8.398 ms & 98.44 $\pm$ 12.64 ms & 109.3 $\pm$ 0.8464 ms \\ 
  nbody & 201.6 $\pm$ 39.57 ms & 100 $\pm$ 29.88 ms & 109.4 $\pm$ 0.8818 ms \\ 
  pystone & 315.1 $\pm$ 112.1 ms & 89.1 $\pm$ 24.46 ms & 109.5 $\pm$ 1.414 ms \\ 
  raytrace-simple & 237.2 $\pm$ 44.22 ms & 98.84 $\pm$ 15.28 ms & 109.4 $\pm$ 0.6615 ms \\ 
  richards & 214.2 $\pm$ 54.36 ms & 95.57 $\pm$ 23.14 ms & 109.5 $\pm$ 1.152 ms \\ 
  spectral-norm & 182.4 $\pm$ 62.91 ms & 77.58 $\pm$ 27.61 ms & 109.7 $\pm$ 1.249 ms \\ 
    \hline
\end{tabular}
}
\endgroup
\caption{A comparison of the three time estimation strategies on a subset of Python benchmarks.}
\label{fig:variance}
\end{figure}
  
To support long-running computations, \system{} instruments $p$ such that every
function and loop calls the \lstinline|maySuspend| function (\cref{f:suspend}),
which may interrupt the computation (by capturing its continuation with
\lstinline|control|) and schedule it
for resumption (using \lstinline|defer|.
The parameter $\delta$ determines how frequently these
interrupts occur.
These interruptions give the browser an opportunity to process
other events, which may include, for example, a user's click on a ``Pause'' button.
To support pausing, \lstinline|maySuspend|
checks if the \lstinline|mustPause| flag is set and calls the \lstinline|onPause|
callback (from the \lstinline|pause| method in \cref{f:stopify-api}) instead of restoring the saved continuation.

The \lstinline|defer| function calls \lstinline|estimateElapsed| to
estimate how much time has elapsed since the last interruption.
To do so, \lstinline|estimateTime| 
counts the number of times it is called
(\lstinline|distance|)
and maintains an estimate of the rate at which it is called
(\lstinline|velocity|). The parameter $t$ determines how frequently
the function actually checks the system time and thus the accuracy of
the estimate. If the true function-call rate is
$v'$, then the estimated time
will be off by a factor of $\frac{\texttt{velocity}}{v'}$, until we resample the system time.

Our approach is significantly less expensive that Skulpt's
approach---which is to report the exact time elapsed---and more accurate than
Pyret's approach---which is to assume a fixed execution rate for all programs.
Since our approach relies on
sampling, it does lose precision. The table in \cref{fig:variance} applies
all three techniques to a subset of our benchmarks and reports
the mean interrupt interval and its standard deviation for each case.
In one particularly bad example, the mean interrupt interval ($\mu$) is $108.3$ms,
with standard deviation ($\sigma$) $88.91$ms. However, 
by Chebyshev's
inequality---$\mathrm{Pr}(|X-\mu| \ge k \sigma) \le 1 / k^2$---even
in this case, $95\%$  of interrupt intervals will be less than
$553$ms, which is still low enough for browsers to be responsive.

\subsection{Blocking, Deep Stacks, and Debugging}
\label{s:stack-tricks}

\system{} also allows programs to
directly suspend and resume their execution with an API call.
This allows the runtime system of a programming language to pause the program while it completes a nonblocking operation,
like a network request or an input event, thereby simulating a blocking operation.
In addition, certain browsers (e.g., Firefox and mobile browsers) have very shallow stacks,
which is a problem for recursive programs.
With the \lstinline|'deep'| option for \lstinline|stacks|
(\cref{f:stopify-api}), \system{} can simulate an arbitrary stack depth (up
to heap size). This mode tracks a stack
depth counter that is updated in every function call.
On reaching a predefined limit, the stack is captured with
\kwcontrol{}, and computation resumes with an empty stack (that closes over
the captured one).
This counter needs just one variable, and so has negligible impact on
performance for programs that don't trigger the stack depth counter:
i.e., programs that don't need it hardly pay for it. This feature is important
to languages like Pyret (\cref{s:pyret}), which encourages
functional programming and abstracts low-level details such as the
stack size.

\begin{figure}
\begin{tikzpicture}
  \node{\pgfimage[width=.95\columnwidth]{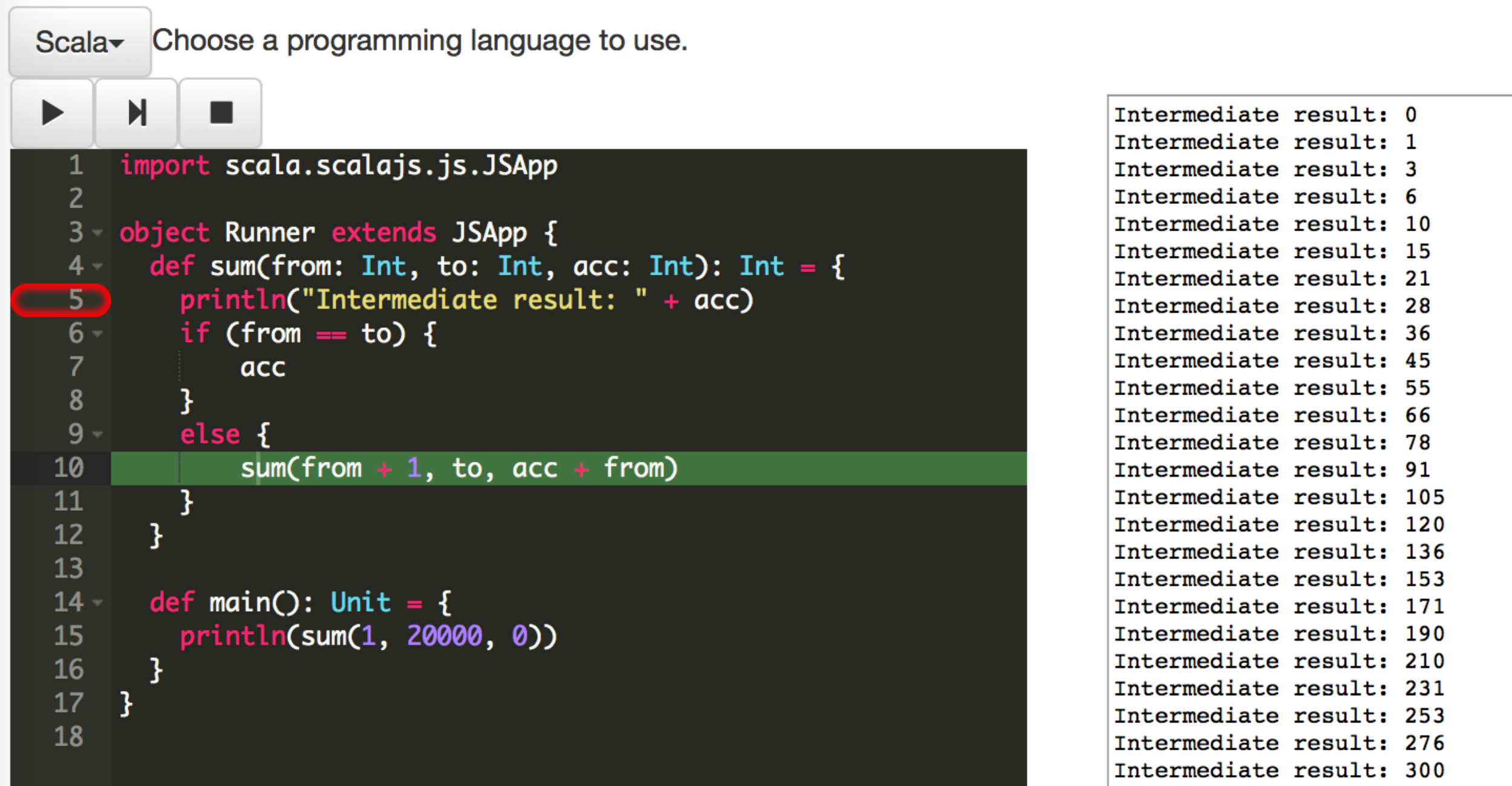}};
\end{tikzpicture}
\caption{Debugging Scala with \system, single-stepped to the
  highlighted line. The red gutter marker
is a breakpoint.}
\label{f:scala-debug}
\end{figure}

Finally, \system{} can be configured to enable breakpoints and stepping.
It does this by instrumenting the program to invoke
\lstinline|maySuspend| before every statement. For breakpoints,
\lstinline|maySuspend| checks if the
currently-paused statement has a breakpoint set. For stepping, we treat the program as if breakpoints are
set on every statement.
\system{} exploits JavaScript \emph{source maps} to allow
breakpoints to be set using locations in the source program. Source
maps are an established technology that is supported by \numstepping{} of the compilers
that we use in our evaluation (\cref{sec:evaluation}). For these languages, \system{} is the
first Web IDE that supports breakpoints and single-stepping, and with no
language-specific effort. (\Cref{f:scala-debug} shows the Web IDE.)

\section{Evaluation}
\label{sec:evaluation}

\begin{figure}
\footnotesize
\begin{tabular}{|l|l|l|}
\hline
Browser & Platform \\
\hline
Google Chrome 60 & Windows 10, Core i5-4690K, 16GB RAM \\
Mozilla Firefox 56 & Windows 10, Core i5-4690K, 16GB RAM \\
Microsoft Edge 40 & Windows 10, Core i5-4690K, 16GB RAM \\
Apple Safari 11 & MacOS 10.13, Core i5-3470S, 8GB RAM \\
ChromeBook & ChromeOS, Celeron N3060, 4GB RAM \\
\hline
\end{tabular}
\caption{The platforms that we use to benchmark \system{}.}
\label{platforms}
\end{figure}

\begin{figure*}
\begin{tikzpicture}
  \node{\pgfimage[width=.95\textwidth]{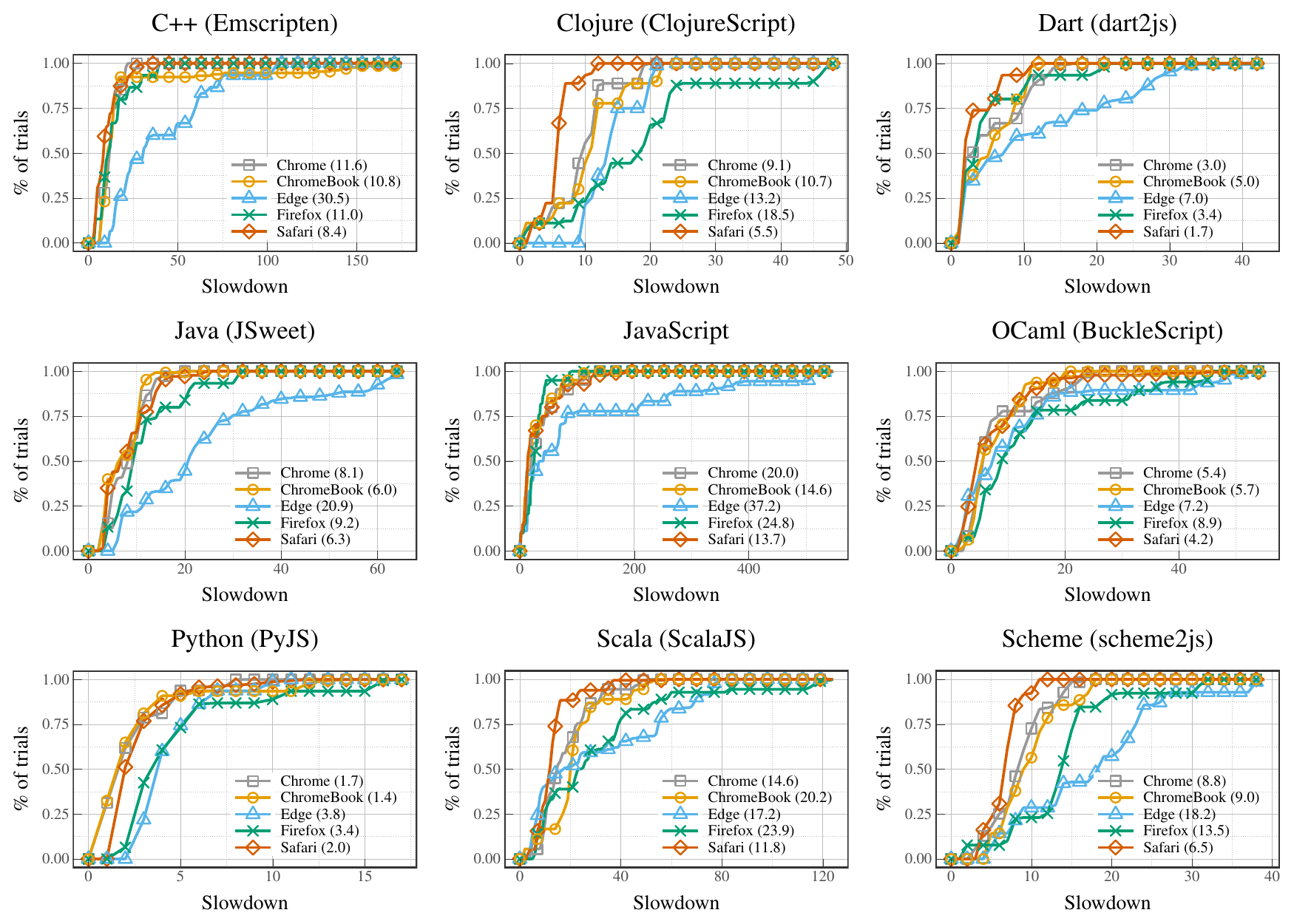}};
\end{tikzpicture}
\caption{CDFs of \system{}'s slowdown on nine languages. The median slowdown is in the legend.}
\label{slowdown-ecdfs}
\end{figure*}

We now evaluate \system{} on a suite of ten languages.
\system{} can introduce an order of magnitude slowdown or more,
but its cost must be understood in context. Without
\system{}, almost all benchmarks either freeze the browser tab or overflow the
JavaScript stack. Furthermore, \system{} supports the first Web IDE
with execution control for
eight of these languages. The two exceptions are Python and Pyret:
we directly compare Stopify to Web IDEs for these languages. This cost may be
insignificant or irrelevant in a Web IDE, and once the code has been developed,
\system{} can be dropped to deploy the full-speed version.

\paragraph{Platform selection} We run all programs on the four major browsers
(Chrome, Firefox, Edge, and Safari) and on a \$200 ChromeBook.
(See \Cref{platforms} for more detailed specifications.)

\subsection{\system{} on Nine Languages}
\label{main-eval}

\begin{figure}
\footnotesize
\begin{tabular}{|l|c|c|}
\hline
\textbf{Platform} & \textbf{Continuations} & \textbf{Constructors} \\
\hline
Edge              & checked-return         & dynamic \\
Safari            & exceptional            & desugar \\
Firefox           & exceptional            & desugar \\
Chrome            & exceptional            & desugar \\
\hline
\end{tabular}
\caption{The best implementation strategy for continuations and constructors
for each browser ($p < 0.01$).}
\label{best-browser-config}
\end{figure}

For our first experiment, we report the cost of \system{} using
the compilers and benchmarks in \cref{compiler-configs}.
(We exclude Pyret here, devoting
\cref{s:pyret} to it.) For eight of these, we make \emph{no
changes to the compiler} and simply apply \system{} to the compiler output.
The only exception is PyJS, which produces JavaScript embedded
in a Web page: we modify the compiler to produce a standalone JavaScript
file. When possible, we use the Computer Language Benchmarks
Game~\cite{benchmarks-game},
which is a collection of toy problems with solutions in several languages,
as our benchmarks. We supplement
the Shootout benchmarks with language-specific benchmarks in some cases.
We only
exclude benchmarks that use language or platform features that the compiler
does not support (e.g., files, foreign function interface, etc.). Therefore,
the excluded benchmarks cannot run in the browser even without \system{}, so
this is not a \system{} limitation.
We run each benchmark ten times, both with and without \system{}.
Finally, we
ensure all benchmarks run for at least two seconds without \system{} by editing
the number of iterations. This ensures that that the benchmark yields control
several times and thus \system{} is properly exercised by every benchmark.
We have \numbench{} benchmarks across all \numlang{} languages.

For each compiler, we configure \system{} to exploit the
sub-language it generates (\cref{compiler-configs}).
\system{} also provides three strategies for implementing continuations
and two strategies for supporting constructors (\cref{s:approach}).
We use microbenchmarks to choose the best settings
for each browser (\cref{best-browser-config}).
Finally, we configure \system{} to yield control every 100 ms, which ensures
that the browser is very responsive. The slowdown that we report
is the ratio of running time with and without \system{}.

We summarize this experiment in \cref{slowdown-ecdfs}, which shows
empirical cumulative distribution functions (CDFs) of the slowdown for each language. In these graphs,
the $x$-axis shows the slowdown and $y$-axis shows the fraction of trials
with slowdown less than $x$. We also report the median slowdown
for each platform in each graph's legend.

\paragraph{Discussion} \Cref{slowdown-ecdfs} shows that (1) there is no best
platform for \system{}, (2) the cost of \system{} depends on the source
language compiler, and (3) the sub-language of JavaScript has a significant
impact on performance.

We find that the slowdown tends to be much lower on Chrome
and Safari than Edge and Firefox. However, we spent months developing \system{}
using Chrome and Safari; thus, we speculate that the slowdowns on Firefox and Edge
can be made much lower. We are pleasantly surprised that the slowdown on our
ChromeBook is comparable to the slowdown on Chrome, despite the fact that the
ChromeBook has far less cache and RAM than the desktop.

\begin{figure*}[!t]
\begin{minipage}[t]{0.30\textwidth}
\centering
\footnotesize
\begin{tabular}{|l|r|r|}
\hline
Benchmark & $\mu$ & 95\% CI \\
\hline
anagram & 0.25 & $\pm$ 0.01 \\
binary-trees & 0.27 & $\pm$ 0.01 \\
fib & 0.25 & $\pm$ 0.00 \\
gcbench & 0.08 & $\pm$ 0.01 \\
nbody & 0.25 & $\pm$ 0.00 \\
pystone & 0.37 & $\pm$ 0.01 \\
schulze & 1.25 & $\pm$ 0.08 \\
spectral-norm & 0.36 & $\pm$ 0.01 \\
\hline
\end{tabular}
\caption{Slowdown relative to Skulpt. (Stopify is faster when $\mu < 1$.) }
\label{fig:skulpt}
\end{minipage}
\qquad
\begin{minipage}[t]{0.30\textwidth}
  \begin{tikzpicture}[baseline=0em]
    \node{\pgfimage[width=0.95\columnwidth]{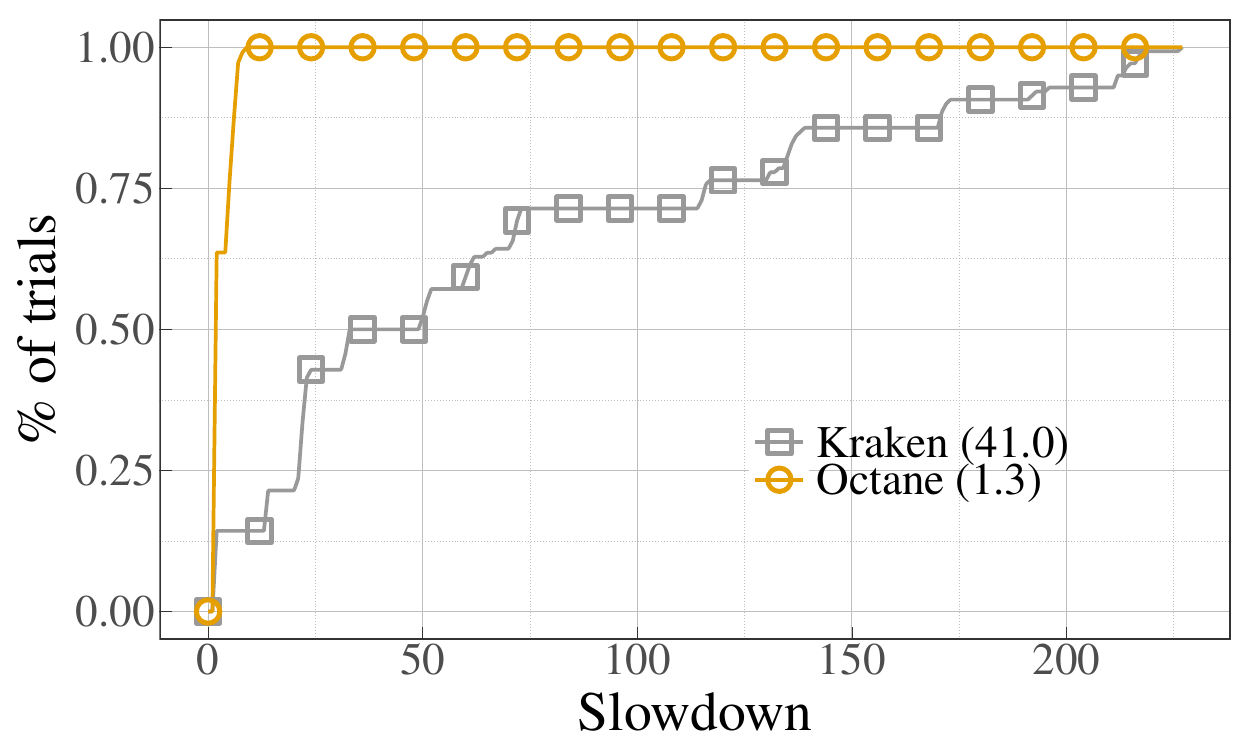}};
  \end{tikzpicture}
  \caption{Slowdown on the Octane and Kraken benchmarks.}
  \label{fig:octane-vs-kraken}
\end{minipage}
\qquad
\begin{minipage}[t]{0.30\textwidth}
  \begin{tikzpicture}[baseline=0em]
    \node{\pgfimage[width=0.95\columnwidth]{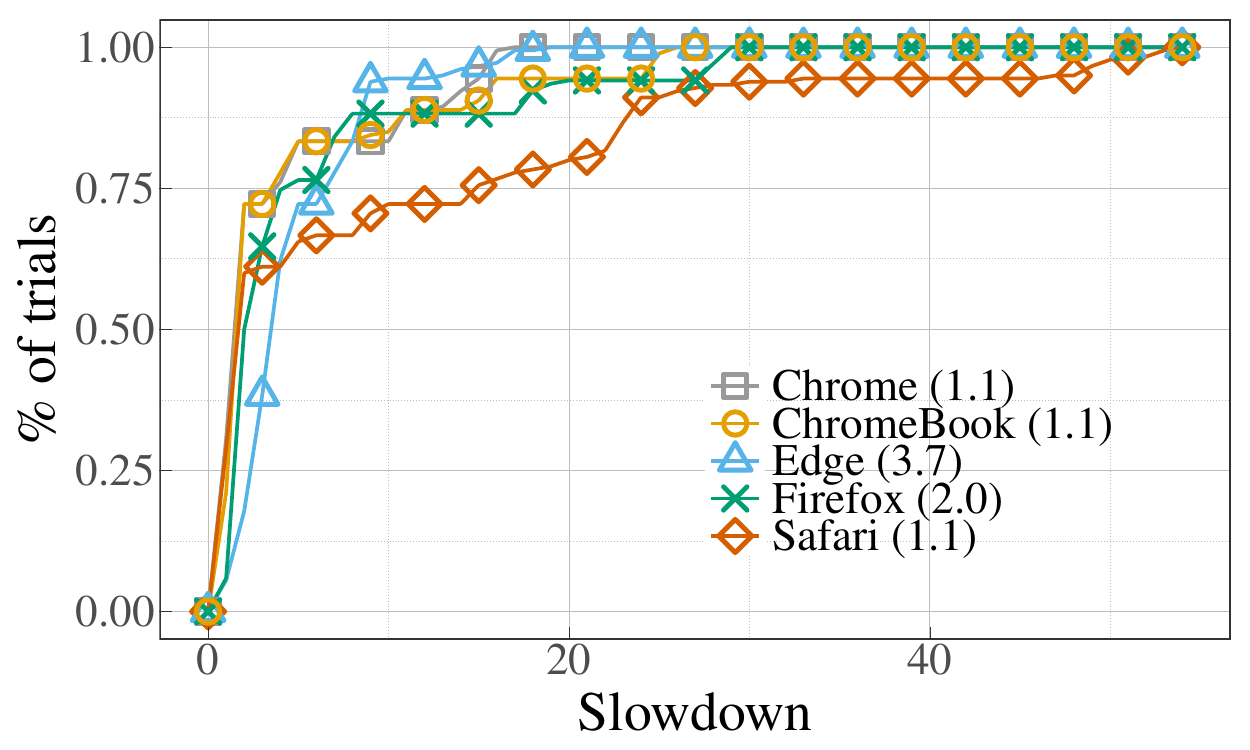}};
  \end{tikzpicture}
  \caption{Slowdown with Pyret.}
  \label{f:pyret-perf}
\end{minipage}
\end{figure*}

We also find that the cost of \system{} varies significantly by source language
compiler. For example, the median slowdown on PyJS ranges from
1.7x---3.8x across all platforms. In contrast, \system{} performs more poorly on
ScalaJS, with slowdowns ranging from 11.8x---23.9x. We attribute these
differences to how these languages are compiled to JavaScript. ScalaJS directly
translates the Scala standard library implementation to JavaScript, instead of mapping Scala
data structures to JavaScript's builtin data structures. (ScalaJS exposes
JavaScript data structures to Scala programs, but our benchmarks do not exploit
them.) PyJS is more lightweight and maps Python data structures to 
JavaScript's builtin data structures. Since \system{} does not change function
signatures, we could improve performance on
ScalaJS by applying \system{} more selectively to the standard
library.

Finally, the cost of \system{} crucially depends on identifying a sub-language
of JavaScript that the compiler produces. The most extreme example is when the
source language is JavaScript itself, so we cannot make any
restrictive assumptions. Implicit operations and getters are the main culprit: the other
nine languages either don't need them or use restricted variants
of these features (\cref{compiler-configs}). We advise compiler
writers who want to leverage \system{} to avoid these features. Fortunately,
existing compilers already do. Still, JavaScript benchmarks that make less use
of these features can avoid the substantial overhead seen in the worst cases.
\Cref{s:octane-kraken} examines JavaScript in more detail.

\paragraph{Effect on Code Size} These benchmarks measure the end-to-end running
time of JavaScript on a Web page, but do not capture page load time. Code size
is a major contributer to delays in load time, especially on mobile platforms.
\system{} increases code-size by a factor of 8x on average with a standard deviation
of 5x.

\paragraph{Native Baselines} The slowdown that we report is the only meaningful
slowdown for platforms that cannot run native code (e.g., ChromeBooks) and
languages that only compile to JavaScript (e.g., Pyret). For other cases, we
report the slowdown incurred by compiling to JavaScript instead of running
natively in \cref{fig:native}. These slowdowns are not
caused by \system{}, but by running code in the browser.

\subsection{Case Study: JavaScript}
\label{s:octane-kraken}

As seen in \cref{main-eval}, \system{} exhibits its worst performance when the
source-language is JavaScript itself. We now compare
the performance of programs compiled
with \system{} across two well-known JavaScript benchmark suites, Octane
\cite{js-octane} and Kraken \cite{moz-kraken}.

\Cref{fig:octane-vs-kraken} shows the slowdowns of each of these benchmark
suites in Chrome 65 when compiled with \system.
We exclude three benchmarks from the Octane benchmark suite due to 1)
limitations of the Babylon parser, 2) the use of event-handlers (which is beyond
the scope of this paper), and 3) the use of non-strict \lstinline|eval| described
in \cref{s:getters-setters}.
We find that the cost of
\system{} differs greatly between these benchmark suites---the median slowdown
across Octane is 1.3x, compared to a median slowdown of 41.0x across Kraken. We
attribute this performance difference to the frequency of implicit calls in
arithmetic operations: \system{} desugars arithmetic to make implicit calls
explicit, and Kraken calls these functions up to an order of magnitude more often
than Octane. This experiment shows that despite heavy
instrumentation, \system{} can achieve low overhead on real JavaScript
applications. \system's performance characteristics are nuanced and largely
dependent on the program being instrumented itself.

\subsection{Case Study: Python (Skulpt)}
\label{s:eval-python}

\begin{figure}
\centering
\footnotesize
\begin{tabular}{|l|r|r|r|r|r|r|}
\hline
Compiler & \multicolumn{3}{|c|}{Chrome} & \multicolumn{3}{|c|}{Firefox} \\
\hline
 & $\mu$ & 95\% CI & Max. & $\mu$ & 95\% CI & Max. \\
\hline
Emscripten & 2.59 & $\pm$ 0.42 & 9.98 & 3.72 & $\pm$ 0.73 & 16.95 \\
ClojureScript & 0.51 & $\pm$ 0.05 & 0.97 & 0.59 & $\pm$ 0.07 & 1.28 \\ 
dart2js & 1.32 & $\pm$ 0.10 & 2.69 & 3.20 & $\pm$ 0.62 & 13.29 \\ 
JSweet & 2.65 & $\pm$ 0.51 & 9.25 & 2.01 & $\pm$ 0.45 & 9.34 \\ 
BuckleScript & 20.48 & $\pm$ 11.80 & 475.89 & 67.98 & $\pm$ 28.96 & 1808.07 \\ 
PyJS & 6.87 & $\pm$ 1.19 & 27.17 & 4.40 & $\pm$ 0.59 & 12.26 \\ 
ScalaJS & 3.58 & $\pm$ 1.14 & 25.92 & 6.30 & $\pm$ 2.79 & 66.17 \\ 
scheme2js & 0.96 & $\pm$ 0.53 & 5.19 & 1.16 & $\pm$ 0.63 & 6.12 \\ 
\hline
\end{tabular}
\caption{Slowdown incurred by compiling to JavaScript (without Stopify) relative to native.}
\label{fig:native}
\end{figure}

In \cref{sec:overview} and \cref{main-eval}, we applied \system{} to
PyJS to get execution control for Python in the browser. We now compare
our approach to Skulpt, which is another Python to JavaScript compiler
that has its own execution control. Skulpt is also widely  used by online Python courses
at
Coursera~\cite{coursera:python-data-structures,coursera:programming-for-everyone},
by Rice University's introductory computer science
courses~\cite{tang:codeskulptor}, by online
textbooks~\cite{miller:how-to-think-like-a-computer-scientist,miller:problem-solving-in-python,guzdial:cs-principals},
and by several schools~\cite{pythonroom,trinket,tynker}. Skulpt can be
configured to either timeout execution after a few seconds or to yield control
at regular intervals, similar to \system{}.
Unfortunately, Skulpt's implementation of continuations is
a new feature that is quite unreliable and often fails to resume execution
after yielding. Therefore, we perform the following experiment that puts \system{} at a
disadvantage: we configure Skulpt to neither yield nor timeout and compare its
performance to \system{} configured to yield every 100 ms.
\Cref{fig:skulpt} shows the normalized runtime of \system{}: a slowdown of 1 is the same as the running time of Skulpt; lower means
\system is faster.
\system{} is substantially faster or competitive with Skulpt on all benchmarks,
despite its handicap in this experiment.

Though both
PyJS and Skulpt are reasonably mature implementations of Python, 
they have their differences: they each seem to pass only portions of
the CPython test suite, and each fail on some benchmarks. Indeed,
Skulpt passes only 8 of our 16
benchmarks. Nevertheless, we believe that this experiment shows that
\system{} is already fast enough to support Python Web IDEs,
while increasing their reliability.

\subsection{Case Study: Pyret}
\label{s:pyret}

\begin{figure}
\begin{subfigure}{\columnwidth}
\begin{lstlisting}
function eachLoop(fun, start, stop) {
  var i = start;
  function restart(_) {
    var res = thisRuntime.nothing;
    if (--thisRuntime.GAS <= 0) { res = thisRuntime.makeCont(); }
    while(!thisRuntime.isContinuation(res)) {
      if (--thisRuntime.RUNGAS <= 0) { res = thisRuntime.makeCont(); }
      else {
        if(i >= stop) {
          ++thisRuntime.GAS;
          return thisRuntime.nothing;
        } else {
          res = fun.app(i);
          i = i + 1;
    } } }
    res.stack[thisRuntime.EXN_STACKHEIGHT++] =
      thisRuntime.makeActivationRecord("eachLoop", restart, true, [], []);
    return res;
  }
  return restart();
}
\end{lstlisting}
\caption{Original, hand-instrumented implementation.}
\label{f:eachLoop-original}
\end{subfigure}

\begin{subfigure}{\columnwidth}  
\begin{lstlisting}
function eachLoop(fun, start, stop) {
  for (var i = start; i < stop; i++) { fun.app(i); }
  return thisRuntime.nothing;
}
\end{lstlisting}
\caption{With \system{}, no instrumentation is necessary.}
\label{f:eachLoop-stopify}
\end{subfigure} 

\caption{A higher-order function from Pyret's standard library that applies
  a function to a sequence of numbers.}
\label{f:eachLoop}
\end{figure}

Pyret~\cite{pyret} is a mostly-functional programming language that
runs entirely in the browser;
it supports proper tail calls, blocking I/O, a
REPL, and interactive animations; it allows users to gracefully
terminate running programs; and it takes care to not freeze the browser tab.
Despite five years of engineering and thousands of users, Pyret still
suffers from issues that produce wrong results or freeze the browser
tab~\cite{pyret-issue-1118, pyret-issue-37, pyret-issue-508,
pyret-issue-1089, pyret-issue-839, pyret-issue-596, pyret-issue-163,
pyret-issue-555, pyret-issue-512, pyret-issue-213, pyret-issue-146,
pyret-issue-145}.

\paragraph{Current Pyret Implementation}

The final phase of the Pyret compiler is a single pass that both (1) translates Pyret
expressions to JavaScript and (2) produces JavaScript that is instrumented to
save and restore the stack. The core Pyret runtime system, which is 6,000 lines of
JavaScript, also serves two roles: it (1) implements Pyret's standard library
and (2) and carefully cooperates with compiled Pyret code.
Since Pyret encourages
functional programming, its standard library has several higher-order
functions, which are implemented in JavaScript to improve performance. These
JavaScript functions are instrumented by hand, since they may appear on
the stack when Pyret needs to pause a user-written function.

Due to this close coupling, the
compiler and runtime system are difficult
to maintain. For example, the Pyret runtime has a function called
\lstinline|eachLoop| that applies a functional argument to a range of
numbers. However, most of its implementation is dedicated to saving and
restoring the stack (\cref{f:eachLoop-original}). In fact, this function has
been rewritten several times to fix bugs in its
instrumentation~\cite{pyret-eachLoop-commit1, pyret-eachLoop-commit2,
pyret-eachLoop-commit3}.

\paragraph{Pyret with \system{}}

We applied \system{} to Pyret, which simplifies both the compiler and runtime
system. However, we also find that \system{} (1)
largely maintains Pyret's performance, (2) supports Pyret's Web IDE, including
animations and the REPL, and (3) exposes new bugs in Pyret.

For the other languages, we use \system{} as a blunt instrument: we leave
an existing compiler and runtime system unchanged and apply \system{} to
the output JavaScript. We apply \system{} more carefully to
Pyret. First, we strip out the portions of the compiler that manage the stack
and use \system{} instead. This shrinks the last phase
of the
compiler from 2,100 LOC to 1,500 LOC (nearly 30\% smaller). Second, 900
LOC of Pyret's runtime system involve stack management. We remove the stack
management logic, which shrinks the code to 350 LOC (60\% smaller), and we
modify Pyret's build process to apply \system{} to this code. For example,
\system{} lets us remove \emph{all} the instrumentation from \lstinline|eachLoop|
(\cref{f:eachLoop}). We envision that compiler authors who choose to use
\system{} themselves will use it in this way to instrument only the necessary
fragments of the language's libraries.

For the other nine languages, we use \system{} to only support long-running
programs, stepping, and graceful termination. However, Pyret requires many
more features (REPL, animations, blocking I/O). All these features
are implemented in JavaScript and use an internal function to pause execution while they
do their work. We made these features work by simply replacing Pyret's
pause function with \system{} APIs.

Finally, \system{} exposed new and nontrivial bugs in Pyret's stack saving
mechanism \cite{pyret-issue-1251}. Two functions in the Pyret
runtime system wrongly assumed that they would never appear on the
stack during continuation capture. By simplifying the runtime system,
\system{} made these errors evident.

\paragraph{Performance} \Cref{f:pyret-perf} compares
Pyret with \system{} to the old
Pyret compiler. On Chrome and Safari, the median slowdown is 1.1x and a number of benchmarks
are \emph{faster} with \system{} than with Pyret' own implementation of
continuations. The median slowdown on Firefox (2.0x) and Edge (3.7x) is
disappointing but, as mentioned in \cref{main-eval}, \system{} is not tuned for
these browsers.

Unfortunately, we also find that some of our benchmarks have more significant
slowdowns (up to 20x). All these benchmarks share the following
attribute: they are deeply recursive programs that require more stack space
than the browser provides. \system{} supports deep stacks
(\cref{s:stack-tricks}), but our implementation performs far more poorly
than Pyret's. We expect to address this problem in the near future:
\system{} should be able to output exactly the kind of code that
Pyret does to implement deep stacks.

\paragraph{Future Work}
There are several more ways to improve our performance on Pyret.
For example, Pyret performs several optimizations in its last phase
that are entangled with its implementation of continuations. We omit these
optimizations in our prototype compiler.
Once ported to \system, they can be applied to other languages too.
Furthermore, there are several ways to simplify Pyret's libraries now that
we have \system{}. For example, many functions use expensive abstractions,
such as callbacks and continuations, to avoid appearing on the stack
during continuation capture. We could rewrite these functions using
loops and apply \system{}, which may improve performance.

\section{Related Work}
\label{relwork}

\begin{figure*}
\begin{subfigure}[b]{0.3\textwidth}
\begin{lstlisting}
j = 0
while (j < 1000000000):
  j = j + 1
\end{lstlisting}
\label{codeschool-infloop}
\caption{CodeSchool (exception).}
\end{subfigure}
\quad
\begin{subfigure}[b]{0.3\textwidth}
\begin{lstlisting}
var i = 0;
while (i++ < 1000000000);
alert(i);
\end{lstlisting}
\label{codepen-infloop}
\caption{CodePen ( wrong result).}
\end{subfigure}
\quad
\begin{subfigure}[b]{0.3\textwidth}
\begin{lstlisting}
import Window
loop : a -> a
loop x = loop x
main = plainText (loop "")
\end{lstlisting}
\label{elm-infloop}
\caption{Elm Debugger (infinite loop).}
\end{subfigure}
\caption{Example programs that show the limitations of several Web IDEs.}
\end{figure*}
  
\paragraph{Web Workers} Web Workers~\cite{web-workers} are essentially isolated
processes for JavaScript and a Web IDE can use them to terminate a user's
program~\cite{ocaml-mooc}.
However, unlike \system, Workers do not
provide richer execution control (e.g., pausing or breakpointing) or deep stacks.
Unlike \system, they also have a limited
interface: they cannot directly access the DOM, and can communicate
only through special shared datatypes~\cite{shared-array-buffer} or by
message
passing.

\paragraph{WebAssembly} WebAssembly~\cite{webassembly} is a new low-level
language in Web browsers.
As of this writing, garbage collection, threads,
tail calls, and host bindings (e.g., access to the DOM) are in the feature
proposal
stage~\cite{webassembly-proposals}. Therefore,
WebAssembly currently does not provide enough features to
prototype a \system{}-like solution, nor are we aware of any
multi-language demonstrations comparable to \cref{sec:evaluation}.
Nevertheless, as it
matures, Web IDEs may want to
switch to it. For that to happen, WebAssembly will need
to support the kind of execution control that \system{} provides
for JavaScript.

\paragraph{Browser Developer Tools} All modern Web browsers include a suite of
developer tools, including a debugger for the JavaScript running in a
page. These tools address the issue of pausing and resuming
execution, and even utilize source maps to allow stepping through execution in a
non-JavaScript source language, but they rely on developer tools being open at
all times. Furthermore, unlike \system, browser developer tools do not enable
long running computations, support arbitrarily deep stacks, or provide a
programming model for synchronous operations atop nonblocking APIs.

\paragraph{Runtime Systems for the Web} There are a handful of compilers that
have features that overlap with \system{}, such as
continuations~\cite{thivierge:gambit-js,whalesong,loitsch:exceptional-continuations},
tail
calls~\cite{thivierge:gambit-js,whalesong,loitsch:exceptional-continuations,pyret},
and graceful termination~\cite{pyret,thivierge:gambit-js,whalesong}.
GopherJS~\cite{gopherjs} supports goroutines using mechanisms related to
\system{} too. These compilers commingle language
translation and working around the platform's limitations.
\system{}
frees compiler authors to focus on language translation.
In addition, \system{} adds continuations to JavaScript itself and supports
features such as exceptional control flow, constructors, ES6 tail calls, and
more. Furthermore, \system{} supports a family of continuation implementation
strategies and we show that the best strategy varies by browser.

Doppio~\cite{vilk:doppio} and Whalesong~\cite{whalesong} implement
bytecode interpreters in the browser that  do not use the JavaScript
stack. Therefore, they can suspend and resume execution.
However, since these are bytecode interpreters for other platforms
(JVM and Racket, respectively), existing compilers and libraries would have to
change significantly to use them.
Browsix~\cite{browsix} acts as
an ``operating system'' for processes in Web Workers. Therefore, it
inherits Web Workers' restrictions: workers cannot share JavaScript values and
cannot interact with the Web page. It also does not provide deep
stacks.
\system{} allows code to run in the main
browser thread, enabling access to the DOM and allowing execution
control for IDEs.

Pivot~\cite{mickens:pivot} isolates untrusted JavaScript into an iframe
and rewrites the program to use generators, which allows blocking I/O between
the iframe and the outside world. \system{} is not an isolation framework
and implements blocking without generators or \textsc{cps} (\cref{sec:callcc}).

\paragraph{Continuations on Uncooperative Platforms}
Many past projects have investigated implementing continuations in
other platforms
that do not natively support them, from C to .NET~\cite{loitsch:exceptional-continuations,tarditi:ml-to-c,cheney-on-the-mta,hieb:rcp,
  narrativejs,jwacs,pettyjohn:cm}. They use a variety of strategies
ranging from \textsc{cps} with trampolines, to C's \lstinline|setjmp| and
\lstinline|longjmp| to effectively provide tail calls.
These systems do not provide \system{}'s other features (\cref{intro}).

Quasar~\cite{quasar} rewrites JVM bytecode to support lightweight threads and
actors, but is quite different from \system{} since JavaScript has no
analogue to low-level JVM instructions that Quasar uses. Kotlin
Coroutines~\cite{kotlin-coroutines} supports asynchronous functions in Kotlin
using a transformation similar to \textsc{cps}. In JavaScript, \system{}'s transformation
performs better than \textsc{cps} and is more compatible with existing code
(\cref{sec:callcc}).

There are a handful of prior implementations of continuations for
JavaScript~\cite{unwinder,debugjs,jwacs}. Unwinder~\cite{unwinder} and
debug.js~\cite{debugjs} use continuations to prototype an in-browser JavaScript
debugger. However, these system support a much smaller fragment of JavaScript
than Stopify. Moreover, they were not designed to support languages that
compile to JavaScript, thus they do not exploit JavaScript sub-languages to
improve performance. We show that deliberately targeting a
sub-language improves performance significantly (\cref{sec:evaluation}).

\paragraph{Other Web IDEs}
\label{s:web-ide-bad-eg}
Codecademy has a Web IDE for JavaScript that does not have a ``stop'' button
and the only way to interrupt an infinite loop is to refresh the page and lose
recent work. CodeSchool has a Web IDE for Python that
imposes a hard timeout on all computations.
example, it cannot run a Python program that counts up to only 1 billion
(\cref{codeschool-infloop}).
Instead, it aborts with the message,
\emph{``TimeoutError: The executor has timed out.''} The same problem affects
Khan Academy's Web IDE for JavaScript~\cite{khan-academy-js}, which
terminates loops with more than five million iterations, printing
\emph{``a while loop is taking too long to run''}.
Codepen is a Web IDE that users to collaboratively develop Web applications
(i.e., \textsc{html}, \textsc{css}, and JavaScript).
The CodePen IDE also terminates long-running
loops, but continues running the program at the next statement after the loop,
thus produces the wrong result (\cref{codepen-infloop}).
Elm~\cite{czaplicki:elm} has a time-traveling
debugger that can step through a program's \textsc{dom} events. However,
infinite loops crash the Elm debugger (\cref{elm-infloop}).
Python Tutor~\cite{codechella} cannot
handle long-running Python programs. If a program
takes too long, it terminates with a message saying that it is not
supported. The Lively debugger~\cite{schuster2012reification}
supports breakpoints and watching variables using an interpreter for a
subset of JavaScript that is written in JavaScript. In contrast, Stopify
compiles JavaScript to JavaScript.

\section{Conclusion}

We have presented \system, a JavaScript-to-JavaScript compiler that
enriches JavaScript with execution control. \system{} allows
Web IDEs to work around the limitations of the JavaScript
execution model. We present a new compilation strategy to support
continuations, identify sub-languages of JavaScript to improve
performance, and improve the responsiveness/performance tradeoff. We
evaluate \system{} by showing that it smoothly composes with
compilers for a diverse set of  \numlang{} programming languages.
\system{} is open source and available at \url{www.stopify.org}.

\section*{Acknowledgments}

We thank Ben Lerner for his contributions to the design of multiple stack
management techniques in Pyret, and for teaching us about them and their tradeoffs;
Emery Berger, John Vilk, and Robert Powers for insightful discussions;
and the anonymous reviewers and our shepherd Christian Wimmer for their
feedback.
This work is supported by the
U.S. National Science Foundation under grants
CCF-1717636. 

\bibliographystyle{ACM-Reference-Format}
\bibliography{main}

\balance

\end{document}